\documentstyle{elsart}
\input{epsf}

\begin{document}
\begin{frontmatter}
  
  \title{ Sources, sinks and wavenumber selection\\ in coupled CGL
    equations\\ and\\ experimental implications for\\ counter-propagating
    wave systems} 
\author[nbi]{Martin van
    Hecke\thanksref{email}}
  \author[rul]{Cornelis Storm} and 
\author[rul]{Wim van Saarloos}
  \address[nbi]{ Center for Chaos and Turbulence Studies, The Niels
  Bohr Institute,\\ Blegdamsvej 17, 2100 Copenhagen \O, Denmark}
  \address[rul]{Instituut--Lorentz, Leiden University, P.O. Box 9506, 2300 RA
  Leiden, \\the Netherlands }
\thanks[email]{Corresponding author. Email: mvhecke@nbi.dk}
\begin{abstract}
  
  We study the coupled complex Ginzburg-Landau (CGL) equations for
  traveling wave systems, and show that sources and sinks are the
  important coherent structures that organize much of the dynamical
  properties of traveling wave systems.
  We focus on the regime in which sources and sinks
  separate patches of left and right-traveling waves, i.e., the case
  that these modes suppress each other.  We present in detail the
  framework to analyze these coherent structures, and show that the
  theory predicts a number of general properties which can be tested
  directly in experiments.  Our counting arguments for the
  multiplicities of these structures show that independently of the
  precise values of the coefficients in the equations, there generally
  exists a symmetric stationary source solution, which sends out waves
  with a unique frequency and wave number.  Sinks, on the other hand,
  occur in two-parameter families, and play an essentially passive
  role, being sandwiched between the sources.  These simple but
  general results imply that sources are important in organizing the
  dynamics of the coupled CGL equations. Simulations show that the
  consequences of the wavenumber selection by the sources is
  reminiscent of a similar selection by spirals in the 2D complex
  Ginzburg Landau equations; sources can send out stable waves,
  convectively unstable waves, or absolutely unstable waves. We show
  that there exists an additional dynamical regime where both single-
  and bimodal states are unstable; the ensuing chaotic states have no
  counterpart in single amplitude equations. A third dynamical
  mechanism is associated with the fact that the width of the sources
  does not show simple scaling with the growth rate $\varepsilon$.
  This is related to the fact that the standard coupled CGL equations
  are {\em not} uniform in $\varepsilon$.  In particular, when the
  group velocity term dominates over the linear growth term, no
  stationary source can exist; however, sources displaying nontrivial
  dynamics can often survive here.  Our results for the existence,
  multiplicity, wavelength selection, dynamics and scaling of sources
  and sinks and the patterns they generate are easily accessible by
  experiments. We therefore advocate a study of the sources and sinks
  as a means to probe traveling wave systems and compare theory and
  experiment. In addition, they bring up a large number of new
  research issues and open problems, which are listed explicitly in
  the concluding section.


\end{abstract}
\end{frontmatter}

\tableofcontents

\section{Introduction}\label{s_intro}

Many spatially extended systems display the formation of patterns when
driven sufficiently far from equilibrium
\cite{newell,ch,newell2,vhhvs,walgraef}.  Examples include convection
\cite{ch}, interfacial growth phenomena \cite{pelce,kassner} like
directional solidification \cite{bechhoefer} and eutectic growth
\cite{faivre}, chemical Turing patterns \cite{ch,walgraef,meron},
the printer instability \cite{printer1,printer2,printer3}, patterns in
liquid crystals \cite{krambuk}, and even biophysical systems
\cite{bio}.  In the typical setup, the homogeneous equilibrium state
turns unstable when a control parameter $R$ (such as the temperature
difference between top and bottom in Rayleigh-B\'enard convection) is
increased beyond a critical value $R_c$.  If the amplitude of the
patterns grows continuously when $R$ is increased beyond $R_c$, the
bifurcation is called supercritical (forward), and a weakly nonlinear
analysis can be performed around the bifurcation point.  A systematic
expansion in the small dimensionless control parameter
$\varepsilon\!:=\! (R-R_c)/R_c$ yields amplitude equations that
describe the slow, large-scale deformations of the basic patterns.

Because near threshold the form of the amplitude or envelope equation
depends mainly on the symmetries and on the nature of the primary
bifurcation (stationary or Hopf, finite wavelength or not, etc.), the
amplitude description has become an important organizing principle of
the theory of non-equilibrium pattern formation. Many qualitative and
quantitative predictions have been successfully confronted with
experiments \cite{ch,newell2,vhhvs,walgraef}.  Even outside their
range of strict applicability, i.e., for finite values of
$\varepsilon$, the amplitude equations are often the simplest
nontrivial models satisfying the symmetries of the underlying physical
system. As such, they can be studied as general models of
nonequilibrium pattern formation.

The most detailed comparison between the predictions of an amplitude
description and experiments has been made \cite{ch} for the type of
systems for which the theory was originally developed \cite{newell},
hydrodynamic systems that bifurcate to a stationary periodic pattern
(critical wavenumber $q_c\neq0$ and critical frequency
$\omega_c\!=\!0$).  The corresponding amplitude equation has real
coefficients and takes the form of a Ginzburg Landau equation; it is
often referred to as the real Ginzburg-Landau equation.  The
coefficients occurring in this equation set length and time scales
only, and for a theoretical analysis of an infinite system, they can
be scaled away. Hence one equation describes a variety of experimental
situations and the theoretical predictions have been compared in
detail with the experimental observations in a number of cases
\cite{ch,newell2,vhhvs,walgraef}.
 
 For traveling wave systems (critical wavenumber $q_c\neq0$ and
 critical frequency $\omega_c\neq0$), there are, however, few examples
 of a direct confrontation between theory and experiment, since the
 qualitative dynamical behavior depends {\em strongly} on the various
 coefficients that enter the resulting amplitude equations\footnote{In
   practice complications may also arise due to the presence of
   additional important slow variables
   \cite{coupling1,coupling2,coupling3,coupling4}.}.  The calculations
 of these coefficients from the underlying equations of motion are
 rather involved and have only been carried out for a limited number
 of systems \cite{bensimon,schopf,kuo,vanhecke,busse}, and in many
 experimental cases the values of these coefficients are not known.  A
 different problem generally arises when dealing with systems of
 counter-propagating waves, where in many cases the standard coupled
 amplitude equations (\ref{coupcgl1},\ref{coupcgl2}) are not uniformly
 valid in $\varepsilon$. 
 Therefore one has to be cautious about the interpretation of results
 based on these equations \cite{cross1,cross2,knobloch,vega2}. We
 return to this issue in section \ref{sss_val}.
 
 It is the main goal of this paper to show that the theory, based on
 the standard coupled amplitude equations
 (\ref{coupcgl1},\ref{coupcgl2}), {\em does} predict a number of
 generic properties of sources and sinks which can be directly tested
 experimentally. In fact, as the results of \cite{alvarez} for
 traveling waves near a heated wire also show, {\em sources} and {\em
   sink} type solutions are the ideal coherent structures to probe the
 applicability of the coupled amplitude equations to experimental
 systems.  The reason is that these coherent structures are, by their
 very nature, based on a competition between left and right-traveling
 waves in the bulk, and, unlike wall or end effects, they do not
 depend sensitively on the experimental details.  Moreover, a study of
 their scaling properties not only yields experimentally testable
 predictions, but also bears on the relation between the averaged
 amplitude equations and the standard amplitude equations (see section
 \ref{sss_val} and \ref{s_sca} below). Finally, as we shall discuss,
 one of our main points is consistent with something which is visible
 in many experiments, namely that the sources determine the wavelength
 in the patches between sources and sinks, and hence organize much of
 the dynamics.
 
 Sources and sinks have been observed in a wide variety of
 experimental systems where oppositely traveling waves suppress each
 other, especially in convection
 \cite{cross1,alvarez,dubois1,dubois2,dubois3,kolss,kolss2,kolss3,daviaud,kapstein1,kapstein2}.
 An example of a one-dimensional source in a chemical system is given
 in \cite{perraud}. To our knowledge, however, they have {\em not}
 been explored systematically in most of these systems. In fact, many
 experimentalists who study traveling wave systems focus on the
 single-mode case --- by perturbing the system or quenching the
 control parameter $\varepsilon$ it is in general possible to
 eliminate the sources and sinks.
 
 Theoretically, some properties of sources and sinks in coupled
 amplitude equations have been analyzed by many workers
 \cite{cross1,cross2,knobloch,vega2,alvarez,joets1,joets2,joets3,coullet3,coullet2,malomed,malomed2,arandw,rovin,neufeld,legavince,manpom}.
 We shall briefly review some of these results in section \ref{ss_his}
 below.  To our knowledge, however, there have been very little
 systematic studies comparing theory and experiment, and we therefore
 advocate a study of these coherent structures as a means to probe
 traveling wave systems.  The two main objectives of this paper are to
 expand the detailed analysis and reasoning underlying the arguments
 of \cite{alvarez}, and to stimulate experimental investigations along
 such lines for other systems as well.
 

\subsection{The coupled complex Ginzburg-Landau equations}\label{ss_ccgle}

When both the critical wavenumber $q_c$ and the critical frequency
$\omega_c$ are nonzero at the pattern forming bifurcation, the primary
modes are traveling waves and the generic amplitude equations are
complex Ginzburg Landau (CGL) equations.  When these primary modes are
essentially one-dimensional and the system possesses left-right
reflection symmetry, the weakly nonlinear patterns are of the form
 \begin{equation}
 \label{convention}
 \mbox{physical fields} \propto A_R e^{-i( \omega_c t - q_c x)}+ A_L e^{-i
   (\omega_c t + q_c x)} + c.c.~,
 \end{equation}
 where $A_R$ and $A_L$ are the complex-valued amplitudes of the right
 and left-traveling waves.  Following arguments from general
 bifurcation theory, i.e., anticipating that these amplitudes are of
 order $\varepsilon^{1/2}$ and that they vary on slow temporal and
 spatial scales, one then finds that the appropriate amplitude
 equations for traveling wave systems with left-right symmetry are the
 coupled CGL equations
 \cite{ch,walgraef,cross1,cross2,knobloch,coullet}
 \begin{eqnarray}
  \partial_{t}A_R + s_0 \partial_{x}A_R & = & \varepsilon A_R
  + (1 + i c_{1}) \partial_{x}^{2}A_R \nonumber\\ & - &  (1 -
  ic_3)|A_R|^2 A_R - g_2 (1 - i c_2) |A_L|^2 A_R ~,\label{coupcgl1} \\
  \partial_{t}A_L  -   s_0 \partial_{x}A_L & = & \varepsilon A_L+
  (1 + i c_{1}) \partial_{x}^{2}A_L \nonumber\\ & - & (1 - i
  c_3)|A_L|^2 A_L - g_2 (1 - i c_2) |A_R|^2 A_L~. \label{coupcgl2}
 \end{eqnarray} 
 In these equations, we have used the freedom to choose appropriate
 units of length, time and of the amplitudes to set various prefactors
 to unity. Our conventions are those of \cite{ch}, except that we
 have, following \cite{cross1}, denoted the coupling coefficient of
 the two modes by $g_2$.  Apart from the ``control parameter''
 $\varepsilon$, there are five important coefficients occurring in
 these equations: $c_1$ and $c_3$ determine the linear and nonlinear
 dispersion of a single mode, $c_2$ determines the dispersive effect
 of one mode on the other, $g_2$ expresses the mutual suppression of
 the modes and $s_0$ is the {\em linear} group velocity of the
 traveling wave modes\footnote{It should be noted that by a rescaling
   one can either fix $\varepsilon$ or $s_0$. Since $\varepsilon$ can
   be varied experimentally, we usually keep $s_0$ at a fixed value
   and vary $\varepsilon$.}. As a function of all these different
 coefficients, many different types of dynamics are found
 \cite{ch,saka1,amen}. 

 
 It is important to stress, following
 \cite{cross1,cross2,knobloch,vega2}, that one has to be cautious
 about the range of validity of the coupled amplitude equations
 (\ref{coupcgl1},\ref{coupcgl2}). When the linear group velocity $s_0$
 is of order $\sqrt{\varepsilon}$, as happens near a co-dimension two
 point in binary mixtures \cite{cross1} or lasers \cite{lasers}, then
 $\varepsilon$ can be removed from the equations by an appropriate
 rescaling of space and time and the amplitude equations are valid
 uniformly in $\varepsilon$. However, in most realistic traveling wave
 systems $s_0$ is of order unity, the amplitude equations do not scale
 uniformly with $\varepsilon$ \cite{cross1}, and their validity is not
 guaranteed.  In practice, the attitude towards this issue has often
 been (either implicitly or explicitly \cite{riecke}) that as they
 respect the proper symmetries, the equations may well yield good
 descriptions of physical systems outside their proper range of
 validity. 
 
 Note in this regard that in a single patch of a left or right
 traveling wave a single amplitude equation for $A_R$ or $A_L$
 suffices; in this case, the linear group velocity term $s_0
 \partial_x A_R $ or $s_0\partial_x A_L$ can be removed by a Galilean
 transformation. The issue of validity of the amplitude equations does
 not arise then (see the discussion in section \ref{sss_pb}), and many
 theoretical studies have focused on this single CGL equation
 \cite{overview1,overview1b,overview1c}.


\subsection{Historical perspective}\label{ss_his}

In this section we will give a brief overview of earlier theoretical work
on sources, sinks and coupled amplitude equations in as far as these
pertain to our work.  It should be noted that grain boundaries for 2D
traveling waves, under the assumption of lateral translational
symmetry, can be described as 1D sources and sinks
\cite{malomed,arandw}; hence some results relevant to the work here
can be found in papers focusing on the 2D case. This explains
the frequent references to early work on grain boundaries in 2D
standing wave patterns \cite{manpom}.  Earlier experimental work
will be discussed in the section on experimental relevance.

\subsubsection{Earlier work on Sources and Sinks}\label{sss_earl}

Early examples of sources and sinks in the literature can be found in
the work by Joets and Ribotta (see \cite{joets1,joets2,joets3} and
references therein), who studied these structures both in experiments
on electroconvection in a nematic liquid crystal, and in simulations
of coupled Ginzburg Landau equations. They focus mainly on nucleation
of sources and sinks, and multiplication processes. Sources and sinks
have also been observed and studied in traveling waves in binary
mixtures \cite{kolss,kolss2,kolss3,kapstein1,kapstein2}. In this system, however, the
transition is weakly subcritical. We will compare some of the results
of these experiments with some of our findings in section \ref{sss_binmix}.

Theoretically, some properties of sources and sinks in coupled
amplitude equations have also been analyzed by Cross \cite{cross1,cross2},
Coullet {\em et al.} \cite{coullet3,coullet2}, Malomed
\cite{malomed,malomed2}, Aranson and Tsimring \cite{arandw} and others
\cite{alvarez,rovin,neufeld}.

Coullet {\em et al.} \cite{coullet3} consider sources and sinks
occurring in one and two-dimen\-sional coupled CGL equations from both a
topological and numerical point of view. In particular, they observe
numerically that patterns in which sources and sinks are present
typically select a unique wavenumber, a feature which plays a central
role in our discussion.

A particular important prediction of Coullet {\em et al.}
\cite{coullet2} was that sources typically exist only a finite
distance above threshold, for $\varepsilon > \varepsilon_{\it\!c}^{\it
  so} > 0$.  
The authors remark that below this threshold, the sources become very
sensitive to noise, and an addition of noise to the coupled CGL
equations was found to inhibit the divergence of sources in this case.
Moreover, they predict that the width of sinks diverges as $ 1/
\varepsilon$ in contrast to what was asserted in \cite{cross1,cross2}
or what was found perturbatively in the limit $s_0 \rightarrow 0$,
$\varepsilon $ finite \cite{malomed}. There appears to have been neither a
systematic numerical check of these predictions nor a comparison with
experiments. In this paper we shall recover the existence of a
critical value $\varepsilon_{\it\!c}^{\it so}$ from a slightly
different angle, and show that $\varepsilon_{\it\!c}^{\it so}$ is only
the critical value above which {\em stationary} source solutions
exist. 
Below $\varepsilon_{\it\!c}^{\it so}$ source-type structures {\em can}
exist, but they are intrinsically dynamical and very large. We
will refer to these structures as {\em non-stationary} sources, as
opposed to the stationary ones we encounter above
$\varepsilon_{\it\!c}^{\it so}$. 
As we will discuss below in section \ref{sss_val}, the prediction of a {\em
  finite} critical value $\varepsilon^{so}_c$ for sources from the
lowest order amplitude equations is a priori questionable, but we
shall argue that the existence of such a critical value is quite
robust for systems where the bifurcation to traveling waves is
supercritical. For systems where the bifurcation in subcritical, there
need not be such a critical value $\varepsilon_{\it\!c}^{so}$. 
This may be the reason that in experiments on traveling waves in
binary fluid convection \cite{kolss}, there does not appear to be
evidence for the nonexistence of stationary sources below a nonzero
value of $\varepsilon^{so}_c$.

Malomed \cite{malomed} studied sources and sinks near the Real
Ginzburg-Landau limit of the coupled CGL equations, and also found
wavenumber selection. Aranson and Tsimring \cite{arandw} considered
domain walls occurring in a 2D version of the complex Swift-Hohenberg
model. Assuming a translational invariance along this domain wall, one
obtains as amplitude equations the coupled 1D CGL equations
(\ref{coupcgl1},\ref{coupcgl2}) with $s_0\!=\!1,c_1 \rightarrow
\infty, c_2\!=\!c_3\!=\!0$ and $g_2\!=\!2$. For that case, a unique
source was found as well as a continuum of sinks.  For the full 2D
problem, a transverse instability typically renders these solutions
unstable.  Finally, Rovinsky {\em et al.} \cite{rovin} studied the
effects of boundaries and pinning on sinks and sources occurring in
coupled CGL equations, and finally we note that some examples of
sources in periodically forced systems are discussed by Lega and Vince
\cite{legavince}.

\subsubsection{Validity of the coupled CGL equations}\label{sss_val}

There is quite some discussion about under what conditions the
standard coupled amplitude equations (\ref{coupcgl1},\ref{coupcgl2})
are valid for counter-propagating wave systems \cite{knobloch,vega2}.
The essential observation is that when $s_0$ is finite, $\varepsilon$
can not be scaled out from the coupled amplitude equations
(\ref{coupcgl1},\ref{coupcgl2}).  


Knobloch and De Luca \cite{knobloch} and Vega and Martel \cite{vega2}
found that under some conditions the amplitude equations for finite
$s_0$ reduce to
\begin{eqnarray}
  \partial_{t}A_R + s_0 \partial_{x}A_R &  = &  \varepsilon A_R
  + (1 + i c_{1}) \partial_{x}^{2}A_R \nonumber \\ & - & (1 -
  ic_3)|A_R|^2 A_R - g_2 (1 - i c_2)< |A_L|^2> A_R ~, \label{aveq1}\\ 
  \partial_{t}A_L -  s_0 \partial_{x}A_L & = & \varepsilon A_L+
  (1 + i c_{1}) \partial_{x}^{2}A_L \nonumber\\ & - & (1 - i
  c_3)|A_L|^2 A_L - g_2 (1 - i c_2) <|A_R|^2> A_L~. \label{aveq2}
\end{eqnarray} 
in the limit $\varepsilon \rightarrow 0$, where $<|A_L|^2>$ and
$<|A_R|^2>$ denote averages
in the co-moving frames of the amplitudes $A_R$ and $A_L$.
Intuitively, the occurrence of the averages stems from the fact that
the group velocity $s_0$ becomes infinite after scaling $\varepsilon$
out of the equations; in other words, when we follow one mode in the
frame moving with the group velocity, the other mode is swept by so
quickly, that only its average value affects the slow dynamics. These
equations have been used in particular to study the effect of boundary
conditions and finite size effects \cite{knobloch,vega2}, but for the
study of sources and sinks they appear less appropriate since they are
effectively decoupled single-mode equations with a renormalized linear
growth term. Nevertheless, we shall see in section \ref{s_sca} that in
the small $\varepsilon$ limit sources and sinks often disappear from
the dynamics, and if so, these equations may yield an appropriate
description of the late-stage regime.

\subsubsection{Complex dynamics in coupled amplitude equations.}\label{sss_cmpl}

In section \ref{s_con} we will discuss chaotic behavior that results
from the source-induced wavenumber selection.  Complex and chaotic
behavior in the coupled amplitude equations has, to the best of our
knowledge,  received very little
attention; notable exceptions are the papers by
Sakaguchi \cite{saka1}, Amengual {\em et al.} \cite{amen} and van
Hecke and Malomed \cite{marmal}.

In the papers of Sakaguchi \cite{saka1}, the coupled CGL equations
(\ref{coupcgl1},\ref{coupcgl2}) were studied in the regime where the
cross-coupling coefficient $g_2$ is close to 1.  It was pointed out
that the transition between single and bimodal states in general
shifts away from $g_2\!=\!1$ when the nonlinear waves show phase or
defect chaos; in some cases this transition can become hysteretic.
Furthermore, periodic states and tightly bound sink/source pairs that
we will encounter in section \ref{ss_bim} below were already obtained
here.

In the recent work by Amengual {\em et al.} \cite{amen}, two coupled
CGL equations with group velocity $s_0$ equal to zero where studied.
The dispersion coefficients $c_1$ and $c_3$ were chosen such that the
uncoupled equations are in the spatio-temporal intermittent regime
\cite{overview1,overview1b,overview1c,homoclons}. Upon increasing the
coupling coefficient $g_2$, sink/source patterns where observed for
$g_2>1$; in these patterns, no intermittency was observed.  We will
comment on this work in section \ref{sss_pb}, and in particular give a
simple explanation of the disappearance of the intermittency.

\subsection{Outline}\label{ss_out}
After discussing the definition of sources and sinks of related
coherent structures in  section \ref{ss_def}, we turn to the 
counting analysis in section \ref{ss_coh}. We focus in our
presentation on the ingredients of the analysis and on the main
results, relegating all technical details of the analysis to
appendices \ref{sss_coh} and \ref{app_coco}.  The essential result is
that one typically finds a unique symmetric source solution with zero
velocity.


We discuss the scaling of the width of sources and sinks with
$\varepsilon$ in section \ref{s_sca}. The main result is that beyond
the critical value $\varepsilon^{\it so}_{\it\!c}$ sources are
intrinsically non-stationary.



In section \ref{s_con}, we discuss the stability of the
waves sent out by the source solutions, and identify three
different mechanisms that may lead to chaotic behavior.
Furthermore we explore numerically some of the richness found
in the coupled amplitude equations. We find a pletora of structures
and possible dynamical regimes. 


Finally, in section \ref{s_dis}, we close our paper by
putting some of our results in perspective, also in relation to the
experiments, and by discussing some
open problems.

\section{Definition of sources and sinks}\label{ss_def}

Sources and sinks arise when the coupling coefficient $g_2$ is
sufficiently large that one mode suppresses the other. Then the system
tends to form domains of either left-moving or right-moving waves,
separated by domain walls or shocks. The distinction between {\em
  sources} or {\em sinks} according to  whether the nonlinear group
velocity points $s$ of the asymptotic plane waves points {\em
  outwards} or {\em inwards} --- see Fig. \ref{f1} --- is crucial
here.   From a 
physical point of view, the group velocity determines the propagation
of small perturbations.  In our definition, a source is an ``active''
coherent structure which sends out waves to both sides, while a sink
is sandwiched between traveling wave states with the group velocity
pointing inwards; perturbations travel away from sources and into
sinks.  Mathematically, it will turn out that the distinction between
sources and sinks in terms of the group velocity $s$ is also precisely
the one that is natural in the context of the counting arguments.

In an actual experiment concerning traveling waves, when one measures
an order parameter and produces space-time plots of its time
evolution, lines of constant intensity indicate lines of constant
phase of the traveling waves (see for example
\cite{alvarez,kolss,kolss2,kolss3}). The direction of the {\em phase
  velocity} $v_{ph}$ of the waves in each single-mode domain is then
immediately clear.
Since $s$ and $v_{ph}$ do not have to have the same sign, one
can not distinguish sources and sinks based on this data alone.
  In passing, we note that it was
found by Alvarez {\em et al.} \cite{alvarez}, and it is also clear
from Fig.\ 11 of \cite{dubois3}, that $v_{ph}$ and $s$ are parallel in
these heated wire experiments, so that the structures which to the
eye look like sources, are {\em indeed} sources according to our
definition.

In the coupled CGL equations (\ref{coupcgl1},\ref{coupcgl2}), $s_0$ is
the {\em linear} group velocity, i.e., the group velocity 
of the fast modes\footnote{We stress that the indices $R$ and $L$ of
  the amplitudes $A_R$ and $A_L$ are associated with the sign of the
  {\em linear group velocity} $s_0$. In writing Eq.
  (\ref{convention}) with $q_c$ and $\omega_c$ positive, we have also
  associated a wave whose phase velocity $v_{ph}$ is to the right with
  $A_R$, and one whose $v_{ph}$ is to the left with $A_L$, but this
  choice is completely arbitrary: At the level of the amplitude
  equations, the sign of the phase velocity of the critical mode plays
  no role.}.
It is important to
realize \cite{physd} that for positive $\varepsilon$, the group
velocity $s$ is {\em different} from $s_0$.  To see this, note that the
coupled CGL equations admit single mode traveling waves of the form
\begin{eqnarray}
A_R & = & a e^{-i(\omega_R t -q x)} ~,~~~ A_L =0~,\\
  & \mbox{or} \nonumber \\
A_L & = & a e^{-i(\omega_L t -q x)} ~,~~~ A_R =0~.
\end{eqnarray}
Substitution of these wave solutions in the amplitude equations 
(\ref{coupcgl1},\ref{coupcgl2}) yields
the nonlinear dispersion relation
\begin{equation}
\omega_{R,L} = \pm s_0 q +  (c_1 + c_3)q^2~,
\end{equation} 
so that the group velocity $s=\partial \omega/\partial q$ of these
traveling waves becomes
\begin{eqnarray}
s_R &=& s_{0,R} + 2 (c_1 + c_3)q~,~~~\mbox{   with }s_{0,R}=s_0~, \label{sdef1}\\
s_L &=& s_{0,L} + 2 (c_1 + c_3)q~,~~~\mbox{   with }s_{0,R}=-s_0~.\label{sdef2}
\end{eqnarray}
When $\varepsilon \downarrow 0$, the band of the allowed $q$
values shrinks to zero, 
and $s$ approaches the linear group velocity $\pm s_0$, as it should.  The
term $2(c_1+c_3)q$ accounts for the change in the group velocity away
from threshold where the total wave number may differ from the
critical value $q_c$. This term involves both the linear and the
nonlinear dispersion coefficient, and its importance increases with
increasing $\varepsilon$. We will therefore sometimes refer to $s$ as
the {\em nonlinear} or {\em total} group velocity, to emphasize the
difference between $s_0$ and $s$.

Clearly it is possible, that $s_0$ and $s$ have opposite signs. Since
the labels $R$ and $L$ of $A_R$ and $A_L$ refer to the signs of {\em
  linear} group velocity $s_0$, if this occurs, the mode $A_R$
corresponds to a wave whose total group velocity $s$ is to the left!
The various possibilities concerning sources and sinks are illustrated
in Fig.\ \ref{f1}.
 
It is important to stress that our analysis focuses on sources and
sinks near the primary supercritical Hopf bifurcation from a
homogeneous state to traveling waves. 
Experimentally, sources and sinks have been studied in detail by
Kolodner \cite{kolss} in his experiments on traveling waves in binary
mixtures. Unfortunately, for this system a direct comparison between
theory and experiments is hindered by the fact that the transition to
traveling waves is {\em subcritical}, not supercritical.

\section{Coherent structures; counting arguments for sources and
  sinks}\label{ss_coh}

\subsection{Counting arguments: general formulations and summary of results}\label{s_cou}

Many patterns that occur in experiments on traveling wave systems or
numerical simulations of the single and coupled CGL equations
(\ref{coupcgl1},\ref{coupcgl2}) exhibit local structures that have an
essentially time-independent shape and propagate with a constant
velocity $v$. For these so-called {\em coherent} structures, the
spatial and temporal degrees of freedom are not independent: apart
from a phase factor, they are stationary in the co-moving frame
$\xi\!=\!x-vt$.  Since the appropriate functions that describe the
profiles of these coherent structures depend only on the single
variable $\xi$, these functions can be determined by ordinary
differential equations (ODE's).  These are obtained by substitution of
the appropriate Ansatz in the original CGL equations, which of course
are partial differential equations.  Since the ODE's can themselves be
written as a set of first order flow equations in a simple phase
space, the coherent structures of the amplitude equations correspond
to certain orbits of these ODE's. Please note that plane waves, since
they have constant profiles, are trivial examples of coherent
structures; in the flow equations they correspond to fixed points.
Sources and sinks connect, asymptotically, plane waves, and so the
corresponding orbits in the ODE's connect fixed points.
Many different coherent structures have been identified within this
framework \cite{homoclons,physd,nozakibekki,exact}.  

The counting arguments that give the multiplicity of such solutions
are essentially based on determining the dimensions of the stable and
unstable manifolds near the fixed points.  These dimensions, together
with the parameters of the Ansatz such as $v$, determine for a certain
orbit the number of constraints and the number of free parameters that
can be varied to fullfill these constraints.  We may illustrate the
theoretical importance of counting arguments by recalling that for the
single CGL equation a continuous family of hole solutions has been
known to exist for some time \cite{nozakibekki}.  Later, however,
counting arguments showed that these source type solutions were on
general grounds expected to come as discrete sets, not as a continuous
one-parameter family \cite{physd}. This suggested that there is some
accidental degeneracy or hidden symmetry in the single CGL equation,
so that by adding a seemingly innocuous perturbation to the CGL
equation, the family of hole solutions should collapse to a discrete
set. This was indeed found to be the case \cite{popp,doelman}.  For
further details of the results and implications of these counting
arguments for coherent structures in the single CGL equation, we refer
to \cite{physd}.

It should be stressed that counting arguments can not prove the
existence  of certain coherent structures, nor can they establish the
dynamical relevance of the solutions. They can
only establish the multiplicity of the solutions, assuming that the
equations have no hidden symmetries. 
Imagine that we know --- either by an explicit construction or from
numerical experiments --- that a certain type of coherent structure
solution does exist. The counting arguments then establish whether
this should be an isolated or discrete solution (at most a member of a
discrete set of them), or a member of a one-parameter family of
solutions, etc. In the case of an isolated solution, 
there are no nearby solutions if we change one of the parameters (like
the velocity $v$) somewhat. For a one-parameter family, 
the counting argument implies that when we start from a known solution
and change the velocity, we have enough other free parameters
available to make sure that there is a perturbed trajectory that flows
into the proper fixed point as $\xi \rightarrow \infty$.

For the two coupled CGL equations (\ref{coupcgl1},\ref{coupcgl2}) the
counting can be performed by a straightforward extension of the
counting for the single CGL equation \cite{physd}.  The Ansatz for
coherent structures of the coupled CGL equations
(\ref{coupcgl1},\ref{coupcgl2}) is the following generalization of the
Ansatz for the single CGL equation:
\begin{equation} 
  A_L (x,t) = e^{-i \omega_L t} \hat{A_L}(x-v t)~,\label{ans2.1}~~~~~~
  A_R (x,t) =  e^{-i \omega_R t} \hat{A_R}(x-vt) ~.
\end{equation}
Note that we take the velocities of the structures in the left and
right mode equal, while the frequencies $\omega$ are allowed to be
different.  This is due to the form of the coupling of the left- and
right-traveling modes, which is through the moduli of the amplitudes.
It obviously does not make sense to choose the velocities of the $A_L$
and $A_R$ differently: for large times the cores of the structures in
$A_L$ and $A_R$ would then get arbitrarily far apart, and at the
technical level, this would be reflected by the fact that with
different velocities we would not obtain simple ODE's for $\hat{A}_L$
and $\hat{A}_R$.  Since the phases of $A_L$ and $A_R$ are not directly
coupled, there is no a priori reason to take the frequencies
$\omega_L$ and $\omega_R$ equal; in fact we will see that in numerical
experiments they are not always equal (see for instance the
simulations presented in Fig. \ref{fdouble}).  Allowing $\omega_L \neq
\omega_R$, the Ansatz (\ref{ans2.1}) clearly has three
free parameters, $\omega_L,\omega_R$ and $v$.

Substitution of the Ansatz (\ref{ans2.1}) into the
coupled CGL equations (\ref{coupcgl1},\ref{coupcgl2}) yields the following
set of ODE's:
\begin{eqnarray}
\partial_{\xi} a_L &=& \kappa_L a_L ~, \label{al}\\
\partial_{\xi} z_L &=& -z_L^2 + \frac{1}{1+ i c_1} 
\left[ -\varepsilon - i \omega_L + (1-i c_3) a_L^2 \right.
  \nonumber\\ & & \left.  + g_2 (1-i c_2) a_R^2 - (v+ s_0) z_L ~
\right],\label{zl}\\ 
\partial_{\xi} a_R &=& \kappa_R a_R~, \label{ar}\\
\partial_{\xi} z_R &=&  -z_R^2 + \frac{1}{1+ i c_1} \left[ -\varepsilon - i 
\omega_R + (1-i c_3) a_R^2 \right. \nonumber \\ & & \left. +  g_2 (1-i
c_2) a_L^2 - (v-s_0) z_R ~\right],\label{zr}
\end{eqnarray}
where we have written
\begin{equation}
\hat{A}_L = a_L e^{i \phi_L}~,~~~
\hat{A}_R = a_R e^{i \phi_R}~.
\end{equation}
and where
$q$, $\kappa$ and $z$ are defined as
\begin{equation} \label{qkap}
  q:=\partial_{\xi} \phi, ~~~ \kappa := (1/a) \partial_{\xi} a~,
  ~~~z:=\partial_\xi \ln ({\hat{A}}) =\kappa+ i q~.
\end{equation} 
Compared to the flow equations for the single CGL equation (see
appendix \ref{sss_coh}), there are two important differences that
should be noted: {\em (i)} Instead of the velocity $v$ we now have
velocities $v\pm s_0$; this is simply due to the fact that the linear
group velocity terms can not be transformed away.  {\em (ii)} The
nonlinear coupling term in the CGL equations shows up only in the flow
equations for the $z$'s.

The fixed points of these flow equations, the points in phase space at
which the right hand sides of Eqs.\ (\ref{al})-(\ref{zr}) vanish,
describe the asymptotic states for $\xi \to \pm \infty$ of the
coherent structures. What are these fixed points? From Eq. (\ref{al})
we find that either $a_L$ or $\kappa_L$ is equal to zero at a fixed
point, and similarly, from Eq.  (\ref{ar}) it follows that either
$a_R$ or $\kappa_R$ vanishes.  For the sources and sinks of
(\ref{coupcgl1},\ref{coupcgl2}) that we wish to study, the asymptotic
states are left- and right-traveling waves.  Therefore the fixed
points of interest to us have either both $a_L$ and $\kappa_R$ 
or both $a_R$ and $\kappa_L$ equal to zero, and we search for
heteroclinic orbits connecting these two fixed points.

As explained before, with counting arguments one determines the
multiplicity of the coherent stuctures 
from {\em (i)} the dimension ${\cal D }_{out}^-$ of the outgoing
(``unstable'') manifold of the fixed point describing the state on the
left ($\xi = -\infty$), {\em (ii)} the dimension ${\cal D}^+_{out}$ of
the outgoing manifold at the fixed point characterizing the state on
the right ($\xi = \infty$) and {\em (iii)} the number ${\cal
  N}_{free}$ of free parameters in the flow equations.  
Note that every flowline of the ODE's corresponds to a particular
coherent solution, with a fully determined spatial profile but with an
{\em arbitrary} position; if we would also specify the point $\xi=0$
on the flowline, the position of the coherent structure would be
fixed.  When we refer to the multiplicity of the coherent solutions,
however, we only care about the profile and not the position.  We
therefore need to count the multiplicity of the {\em orbits}.  In
terms of the quantities given above, one thus expects a $({\cal
  D}^-_{out} -1 - {\cal D}^+_{out} + {\cal N}_{free}) $--parameter
family of solutions; the factor $-1$ is associated with the invariance
of the ODE's with respect to a shift in the pseudo-time $\xi$ which
leaves the flowelines invariant. In terms of the coherent structures,
this symmetry is the translational invariance of the amplitude
equations.

When the number $({\cal D}_{out}^- -1 - {\cal D}^+_{out} + {\cal
  N}_{free})$ is zero, one expects a discrete set of solutions, while
if this number is negative, one expects there to be no solutions at
all, generically.  {\em Proving} the existence of solutions, within
the context of an analysis of this type, amounts to proving that the
outgoing manifold at the $\xi=-\infty$ fixed point and the incoming
manifold at the $\xi=\infty$ fixed point intersect. Such proofs are in
practice far from trivial --- if at all possible --- and will not be
attempted here.
 
Conceptually, counting arguments are simple, since the dimensions
${\cal D}_{out}^-$ and ${\cal D}_{out}^+$ are just determined by
studying the linear flow in the neighborhood of the fixed points.
 Technically, the analysis of the
coupled equations is a straightforward but somewhat involved extension 
of the earlier findings for the single CGL. We therefore prefer to only
quote the main result of the analysis, and to relegate all
technicalities to appendix \ref{app_coco}.

For sources and sinks, always one of the two modes vanishes at the
relevant fixed points. We are especially interested in the case in
which the effective value of  $\varepsilon$, defined as
\begin{equation}
\varepsilon^L_{\it\!eff\!} := \varepsilon- g_2 |a_R|^2 ~,~~~~~
\varepsilon^R_{\it\!eff\!} := \varepsilon- g_2 |a_L|^2~.\label{epseffdef}
\end{equation}
is {\em negative} for the mode which is suppressed. In this case small
perturbations of the suppressed mode decay to zero in each of the
single-amplitude domains, so this situation is then {\em stable}.
E.g., for a stable source configuration as sketched in Fig.\ 
\ref{figuress}, $\varepsilon^R_{\it\!eff\!}$ should be negative on the
left, and $\varepsilon^L_{\it\!eff\!}$ should be negative on the right
of the source. We will focus below on the results for this regime of
full suppression of one mode by the other.

The basic result of our counting analysis for the multiplicity of source and
sink solutions is that when $\varepsilon_{\it\!eff\!}<0$ the counting
arguments for {\em ``normal''} sources and sinks (the linear group
velocity $s_0$ and the nonlinear group velocity $s$ of the
same sign), is simply that
\begin{itemize}
\item {\em Sources occur in discrete sets. Within these sets, as a result
  of the left-right symmetry for $v=0$, we expect a  stationary,
    symmetric source to occur.}
\item {\em Sinks occur in a two parameter family.}
\end{itemize}
Notice that apart from the conditions formulated
above, these findings are completely independent of the precise values 
of the coefficients of the equations. This gives these results their
predictive power. Essentially all of the results of the remainder of
this paper are based on the first finding that sources come in
discrete sets, so that they fix the properties of the states
in the domains they separate.

As discussed in Appendix \ref{app_coco} the multiplicity of {\em
  anomalous} sources is the same as for normal sources and sinks in
large parts of parameter space, but larger multiplicities {\em can}
occur. Likewise, sources with $\varepsilon_{\it\!eff\!}>0 $ may occur
as a two-parameter family, although most of these are expected to be
unstable
(Appendix \ref{esoteric}).  We shall see in Section \ref{s_con} that
in this case, which happens especially when $g_2$ is only slightly
larger than 1, new nontrivial dynamics can occur.
 
\subsection{Comparison between shooting and direct simulations}
\label{ss_compare}
Clearly, the coherent structure solutions are by construction {\em
  special} solutions of the original partial differential equations.
The question then arises whether these solutions are also dynamically
relevant, in other words, whether they emerge naturally in the long
time dynamics of the CGL equation or as ``nearby'' transient solutions
in nontrivial dynamical regimes. For the single CGL equation, this has
indeed been found to be the case
\cite{homoclons,physd,thualfauve,fauvethual,hakim,elphick}.
To check that this is also the case here, we have performed simulations of the
coupled CGL equations and compared the sinks and sources that are found
there to the ones obtained from the ODE's (\ref{al}-\ref{zr}).  Direct
integration of the coupled CGL equations was done using a
pseudo-spectral code.  The profiles of uniformly translating coherent
structures where obtained by direct integration of the ODE's
(\ref{al}-\ref{zr}), shooting from both the $\xi\!=\!+\infty$ and
$\xi\!=\!-\infty$ fixed points and matching in the middle.

In Fig. \ref{f3}a, we show a space-time plot of the evolution towards
sources and sinks, starting from random initial conditions. The grey
shading is such that patches of $A_R$ mode are light and $A_L$ mode
are dark.  Clearly, after a quite short transient regime, a stationary
sink/source pattern emerges.  In Fig. \ref{f3}b we show the amplitude
profiles of $|A_R|$ (thin curve) and $|A_L|$ (thick curve) in the
final state of the simulations that are shown in Fig.  \ref{f3}a.  In
Fig. \ref{f3}c and d we compare the amplitude and wavenumber profile
of the source obtained from the CGL equations around $x=440$ (boxes)
to the source that is obtained from the ODE's (\ref{al}-\ref{zr})
(full lines).  The fit is excellent, which illustrates our finding
that sources are stable and stationary 
in large regions of parameter space 
and that their profile is
completely determined by the ODE's associated with the Ansatz
(\ref{ans2.1}).

However, the CGL equations posses a large number of coefficients that
can be varied, and it will turn out that there are several mechanisms
that can render sources and source/sink patterns unstable. We will
encounter these scenarios in sections \ref{s_sca} and 
\ref{s_con}.

\subsection{Multiple discrete sources}\label{eso_mult}

As we already pointed out before, the fact that sources come in a
discrete set does not imply that there exists only one unique source
solution.  There could in principle be more solutions, since the
counting only tells us that infinitesimally close to any given
solution, there will not be another one.

Fig. \ref{fdouble} shows an example of the occurrence of two different
isolated source solutions. The figure is a space-time plot of a
simulation where we obtained two different sources, one of which is an
anomalous one ($s$ and $s_0$ of opposite sign). One clearly sees the
different wavenumbers emitted by the two structures, and sandwiched in
between these two sources is a single amplitude sink, whose velocity
is determined by the difference in incoming wavenumbers. We have
checked that the wavenumber selected by the anomalous source is such
that the counting still yields a discrete set. If we follow the
spatio-temporal evolution of this particular configuration, we find
highly nontrivial behavior which we do not fully understand as of yet
(not shown in Fig.  \ref{fdouble}).

These findings illustrate our belief that the ''normal'' sources and
sinks are the most relevant structures one expects to encounter.  It
therefore appears to be safe to ignore the possible dynamical
consequences of the more esoteric structures, which one {\em a priori}
cannot rule out. The main complication of the possible occurrence of
multiple discrete sources, as in Fig. \ref{fdouble}, is that single
amplitude sinks can arise in the patches separating them.  The motion
of these sinks can dominate the dynamics for an appreciable time.

\section{Scaling properties of sources and sinks for
small $\varepsilon$}\label{s_sca}

In this section we study the scaling properties and dynamical behavior
of sources and sinks in the limit where $\varepsilon$ is small.
This is a nontrivial issue, since due to the presence of the linear
group velocity $s_0$, the coupled CGL equations do not scale uniformly
with $\varepsilon$. We focus in particular on the width of the sources
and sinks.  The results we obtain are open for experimental testing,
since the control parameter $\varepsilon$ can usually be varied quite
easily.  The behavior of the sources is the most interesting, and we
will discuss this in sections \ref{ss_san} and \ref{ss_num}.  Using
arguments from the theory of front propagation, we recover the result
from Coullet {\em et al.} \cite{coullet2} that there is a finite
threshold value for $\varepsilon$, below which no {\em coherent}
sources exist (section \ref{ss_san}).  For $\varepsilon $ below this
critical value, there are, depending on the initial conditions,
roughly two different possibilities.  For well-separated sink/source
patterns, we find {\em non-stationary} sources whose average width
scales as $1/\varepsilon$ (in possible agreement with the experiments
of Vince and Dubois \cite{dubois2}; see section \ref{sss_hw}). These
sources can exist for arbitrarily small values of $\varepsilon$.  For
patterns with less-well separated sources and sinks, we typically find
that the sources and sinks annihilate each other and disappear
altogether. The system evolves then to a single mode state, as
described by the averaged amplitude equations equations
(\ref{aveq1}-\ref{aveq2}).  These scenarios are discussed in section
\ref{ss_num} below.  By some simple analytical arguments we obtain
that the width of coherent sinks diverges as $1/\varepsilon$;
typically these structures remain stationary (see section
\ref{ss_sink}).

\subsection{Coherent sources: analytical arguments}\label{ss_san}

By balancing the linear group velocity term with the second order
spatial derivate terms, we see that the coupled amplitude equations
(\ref{coupcgl1}-\ref{coupcgl2}) may contain solutions whose widths
approach a finite value of order $1/s_0$ as $\varepsilon \rightarrow
0$.  As pointed out in particular by Cross \cite{cross1,cross2}, this
behavior might be expected near end walls in finite systems; in principle,
it could also occur for coherent structures such as sources and sinks
which connect two oppositely traveling waves.  Solutions of this type
are {\em not} consistent with the usual assumption of separation of
scales (length scale $\sim \varepsilon^{-1/2}$) which underlies the
derivation of amplitude equations. One should interpret the results
for such solutions with caution.

As we shall discuss below, the existence of stationary, coherent
sources is governed by a finite critical value $\varepsilon^{so}_c$,
first identified by Coullet {\em et al.} \cite{coullet2}.  Since the
coupled amplitude equations (\ref{coupcgl1}-\ref{coupcgl2}) are only
valid to lowest order in $\varepsilon$, the question then arises
whether the existence of this finite critical values
$\varepsilon^{so}_c$ is a peculiarity of the lowest order amplitude
equations.
Since this threshold is determined by the interplay of the linear
group velocity and a front velocity, which are both defined for
arbitrary $\varepsilon $, we will argue that the existence of a
threshold is a robust property indeed.

We now proceed by deriving this critical value $\varepsilon_{\it\!c
  }^{\it so}$ from a slightly different perspective than the one that
underlies the analysis of Coullet {\em et al.} \cite{coullet2}, by
viewing wide sources as weakly bound states of two widely separated
fronts. Indeed, consider a sufficiently wide source like the one
sketched in Fig. \ref{wide}a in which there is quite a large interval
where both amplitudes are close to zero\footnote{It is not completely
  obvious that wide sources necessarily have such a large zero patch,
  but this is what we have found from numerical simulations.  Wide
  sinks actually will turn out not to have this property.}.
Intuitively, we can view such a source as a weakly bound state of two
fronts, since in the region where one of the amplitudes crosses over
from nearly zero to some value of order unity, the other mode is
nearly zero. Hence as a first approximation in describing the fronts
that build up the wide source of the type sketched in Fig.
\ref{wide}a, we can neglect the coupling term proportional to $g_2$ in
the core-region. The resulting fronts will now be analyzed in the
context of the single CGL equation.

Let us look at the motion of the $A_R$ front on the right (by symmetry
the $A_L$ front travels in the opposite direction). As argued above,
its motion is governed by the single CGL equation in a frame moving
with velocity $s_0$
\begin{equation}\label{scgl}
  (\partial_t+s_0\partial_x) A_R = \varepsilon A_R +
  (1+ic_1)\partial_x^2 A_R -(1-ic_3)|A_R|^2 A_R~.
\end{equation}
The front that we are interested in here
corresponds to a front propagating ''upstream'', i.e., to the left,
into the {\em unstable} $A_R\!=\!0$ state. Such fronts have been
studied in detail \cite{physd}, both in general and for the
single CGL equation specifically.

Fronts propagating into unstable states come in two classes, depending
on the nonlinearities involved. Typically, when the nonlinearities are
saturating, as in the cubic CGL equation (\ref{scgl}), the asymptotic
front velocity $v_{\it front}$ equals the {\em linear spreading
  velocity} $v^*$. This $v^*$ is the velocity at which a small
perturbation around the unstable state grows and spreads according to
the {\em linearized} equations. For Eq. (\ref{scgl}), the velocity
$v^*$ of the front, propagating into the unstable $A=0$ state, is given
by \cite{physd}
\begin{equation}\label{vstar}
v^*=s_0-2 \sqrt{\varepsilon(1+c_1^2)}~.
\end{equation}
The parameter regime in which the selected front velocity is $v^*$ is
often referred to as the ``linear marginal stability''
\cite{lmsc,lmsc2} or ``pulled fronts''
\cite{stokes,goldenfeld,ebert} regime, as in this regime the
front is ''pulled along'' by the growing and spreading of linear
perturbations in the tip of the front.

For small $\varepsilon$, the velocity $v^*\!=\!v_{\it front}$ is
positive, implying that the front moves to the right, while for large
$\varepsilon$, $v^*$ is negative so that the front moves to the left.
Intuitively, it is quite clear that the value of $\varepsilon$ where
$v^*\!=\!0$ will be an important critical value for the dynamics,
since for larger $\varepsilon$ the two fronts sketched in Fig.
\ref{wide}a will move towards each other, and some kind of source
structure is bound to emerge.  For
$\varepsilon<\varepsilon_{\it\!c}^{\it so}$, however, there is a
possibility that a source splits up into two retracting fronts.  Hence
the critical value of $\varepsilon$ is defined through
$v^*(\varepsilon_{c}^{\it so})=0$, which, according to Eq.
(\ref{vstar}) yields
\begin{equation}
\varepsilon_c^{\it so}=s_0^2/(4+ 4 c_1^2)~.
\end{equation}
We will indeed find that the width of {\em coherent} sources diverges
for this value of $\varepsilon$; however, the sources will not
disappear altogether, but are replaced by {\em non-stationary} sources
which can not be described by the coherent structures Ansatz
(\ref{ans2.1}).

\subsection{Sources: numerical simulations}\label{ss_num}

By using the shooting method, i.e., numerical integration of the ODE's
(\ref{al}-\ref{zr}), to obtain coherent sources, we have studied the
width of the coherent sources as a function of $\varepsilon$.  The
width is defined here as the distance between the two points where the
left- and right traveling amplitudes reach 50 \% of their respective
asymptotic values.  In Fig. \ref{wide}b, we show how the width of
coherent sources varies with $\varepsilon$.  For the particular choice
of coefficients here ($c1\!=\!c_3\!=\!0.5, c_2\!=\!0,g_2\!=\!2$ and
$s_0\!=\!1$), $\varepsilon_c^{\it so}\!=\!0.2$, and it is clear from this
figure, that the width of stationary source solutions of Eqs.
(\ref{scgl}) diverges at this critical value\footnote{ Note that by a
  rescaling of the CGL equations, one can set $s_0\!=\!1$ without loss of
  generality.}.

In dynamical simulations of the full coupled CGL equations however,
this divergence is cut off by a crossover to the dynamical regime
characteristic of the $\varepsilon<\varepsilon_{\it\!c}^{\it so}$
behavior.  Fig. \ref{wide}c is a space-time plot of $|A_L|+|A_R|$ that
illustrates the incoherent dynamics we observe for
$\varepsilon<\varepsilon_{\it\!c}^{\it so}$.  The initial condition
here is source-like, albeit with a very small width.  In the
simulation shown, we see the initial source flank diverge as we would
expect since $s_0\!>\! v^*$. As time progresses, right ahead of the
front a small 'bump' appears: as we mentioned before, both amplitudes
are to a very good approximation zero in that region, so the state
there is unstable (remember that though small, $\varepsilon$ is still
nonzero). This bump will therefore start to grow, and will be advected
in the direction of the flank. The flank and bump merge then and the
flank jumps forward.  The average front velocity is thus enhanced.
The front then slowly retracts again, and the process is repeated,
resulting in a ``breathing'' type of motion. For longer times these
oscillations become very, very small.  For this particular choice of
parameters, they become almost invisible after times of the order
3000; however, a close inspection of the data yields that the sources
never become stationary but keep performing irregular oscillations.
Since these fluctuations are so small, it is very likely that to an
experimentalist such sources appear to be completely stationary.

From the point of view of the stability of sources, we can think of
the change of behavior of the sources as a core-instability. This
instability is basically triggered by the fact that wide sources have a
large core where both $A_L$ and $A_R$ are small, and since the neutral
state is unstable, this renders the sources unstable.  The difference
between the critical value of $\varepsilon$ where the instability sets
in and $\varepsilon_{\it\!c}^{\it so}$ is minute, and we will not
dwell on the distinction between the two.\footnote{For a similar
  scenario in the context of non-homogeneously coupled CGL equations,
  see \cite{marmal}.} Although all our numerical results are in accord
with this scenario, one should be aware, however, that it is not
excluded that other types of core-instabilities exist is some regions
of parameter space\footnote{ An example of a similar scenario is
  provided by pulses in the single quintic CGL equation. Pulses are
  structures consisting of localized regions where $|A|\neq0$. The
  existence and stability of pulse solutions can, to a large extent, be
  understood by thinking of a pulse as a bound state of two fronts
  \cite{physd}. However, recent perturbative calculations near the
  non-dissipative (Schr\"odinger-like) limit \cite{kaup} have shown
  that in some parameter regimes a pulse can become unstable
against a localized mode. This particular instability can not simply be
understood by viewing a pulse as a bound state of two fronts.}.
Furthermore, it should be pointed out that when $\varepsilon$ is
below $\varepsilon_{\it\!c}^{\it so}$, there is {\em no} stationary
albeit unstable source! The dynamical sources can than {\em not} be viewed
as oscillating around an unstable stationary source.

The weak fluctuations of the source flanks are very similar to the
fluctuations of domain walls between single and bimodal states in
inhomogeneously coupled CGL equations as studied in \cite{marmal}.
Completely analogous to what is found here, there is a threshold given
in terms of $\varepsilon$ and $s_0$ for the existence of stationary
domain walls, which we understand now to result from a similar
competition between fronts and linear group velocities.  Beyond the
threshold, dynamical behavior was shown to set in, which, depending on
the coefficients, can take qualitatively different forms; similar scenarios
can be obtained for the sources here.

The main ingredient that generates the dynamics seems to be the
following. For a very wide source, we can think of the flank of the
source as an isolated front.  However, the {\em tip} of this front
will always feel the other mode, and it is precisely this tip which
plays an essential role in the propagation of ``pulled'' fronts
\cite{lmsc,lmsc2,stokes,goldenfeld,ebert}! Close inspection
of the numerics shows that near the crossover between the front regime
and the interaction regime, oscillations, phase slips or kinks are
generated, which are subsequently advected in the direction of the
flank. These perturbations are a {\em deterministic} source of
perturbations, and it is these perturbations that make the flank jump
forward, effectively narrowing down the source.

The jumping forward of the flank of the source for $\varepsilon$ just
below $\varepsilon_{\it\!c}^{\it so}$ is reminiscent to the mechanism
through which traveling pulses were found to acquire incoherent
dynamical behavior, if their velocity was different from the linear
group velocity \cite{theotherprl}.  In extensions of the CGL equation,
it was found that if a pulse would travel slower than the linear
spreading speed $v^*$, fluctuations in the region just ahead of the
pulse could grow out and make the pulse at one point ''jump ahead''.
In much the same way  the fronts can be viewed to ''jump ahead''
in the wide source-type structures below $\varepsilon_{\it\!c}^{\it so}$
 when the fluctuations ahead of it grow
sufficiently large.

In passing, we point out that we believe
these various types of ``breathing dynamics'' to be a general feature
of the interaction between local structures and fronts.  Apart from
the examples mentioned above, a well known example of incoherent
local structures are the oscillating pulses observed by Brand and
Deissler in the quintic CGL \cite{brand}.
Also in this case we have
found that these oscillations are due to the interaction with a front,
but instead of a pulled front it is a {\em pushed} front that drives the
oscillations here \cite{ewald}.

Returning to the discussion of the behavior of the wide
non-stationary sources, we show in Fig. \ref{wide}d   the
(inverse) average width of the dynamical sources  for small
$\varepsilon$. These simulations where done in a large system (size
2048), with just one source and, due to the periodic boundary
conditions, one sink. If one slowly decreases $\varepsilon$, one finds
that the average width of the sources diverges roughly as
$\varepsilon^{-1}$ (see the inset of Fig.  \ref{wide}d). However, if
one does not take such a large system, i.e., sources and sinks are not
so well separated, we often observed that, after a few oscillations of
the sources, they interact with the sinks and annihilate. In many
case, especially for small enough $\varepsilon$, {\em all} sources and
sinks disappear from the system, and one ends up with a state of only
right or left traveling wave.  Since no sources or sinks can occur in
the average equations (\ref{aveq1},\ref{aveq2}), this behavior seems
precisely to be what these average equations predict.  In a sense,
this regime without sources and sinks follows nicely from the ordinary
CGL equations when $\varepsilon \downarrow 0$.

In conclusion,  we arrive at the following scenario.

\begin{itemize}
  
\item For $\varepsilon>\varepsilon_{\it\!c}^{\it so}$, sources are
   {\em stationary} and stable, provided that the waves they
  send out are stable. The structure of these stationary source
  solutions is given by the ODE's (\ref{al}-\ref{zr}), and their
  multiplicity is determined by the counting arguments.
    
\item When $\varepsilon \downarrow \varepsilon_{\it\!c}^{\it so}$, the
  source width rapidly increases,
and for
  $\varepsilon=\varepsilon_{\it\!c}^{\it so}$, the size of the coherent
  sources (i.e., solutions of the ODE's (\ref{al}-\ref{zr}))
  diverges, in agreement with the picture of a
  source consisting of two weakly bound fronts. 
 For a value of $\varepsilon$ just above $\varepsilon_{\it\!
    c}^{\it so}$, the sources have a wide core where both $A_R$ and
  $A_R$ are close to zero, and   these sources turn unstable. Our
  scenario is that in this regime a source consists essentially of two 
of the ``nonlinear global modes'' of Couairon and Chomaz
\cite{couairon}. Possibly, their analysis can be extended to study
the divergence of the source width as $\varepsilon \downarrow
\varepsilon_c^{so}$.

\item For $\varepsilon<\varepsilon_{\it\!c}^{\it so}$, {\em wide},
  {\em non-stationary} sources can exist. Their dynamical behavior is
  governed by the continuous emergence and growth of fluctuations in
  the region where both amplitudes are small, resulting in an
  incoherent ``breathing'' appearance of the source.  For long times,
  these oscillations may become very mild, especially when
  $\varepsilon$ is not very far below $\varepsilon_{\it\!c}^{\it so}$.
  
\item In the limit for $\varepsilon \downarrow 0$, there are,
  depending on the initial conditions, two possibilities. For random
  initial conditions, pairs of sources and sinks annihilate and the
  system often ends up in a single mode state, which is consistent
  with the 'averaged equation' picture discussed in section \ref{sss_val}. This happens in particular in
  sufficiently small systems.  Alternatively, in large systems, one
  may generate well-separated sources and sinks. In this case the
  average width of the incoherent sources diverges as $1/\varepsilon$,
  in apparent agreement with the experiments of Vince and Dubois
  \cite{dubois2} (see section \ref{sss_hw} for further discussion of
  this point).

\end{itemize}

We finally note that our discussion above was based on the fact that
near a supercritical bifurcation, fronts propagating into an unstable
state are ''pulled'' \cite{stokes,goldenfeld,ebert} or
''linear marginal stability'' \cite{lmsc,lmsc2} fronts: $v_{\it
  front}\!=\!v^*$.  It is well-known that when some of the nonlinear
terms tend to enhance the growth of the amplitude, the front velocity
can be higher: $v_{\it front}>v^*$
\cite{lmsc,lmsc2,stokes,goldenfeld,ebert}.
These fronts, which occur in particular near a subcritical
bifurcation, are sometimes called ''pushed''
\cite{stokes,goldenfeld,ebert} or ''nonlinearly marginal
stability'' \cite{physd,lmsc2} fronts. In this case it can happen that
the front velocity remains large enough for stable stationary sources
to exist all the way down to $\varepsilon\!=\!0$. We believe that this
is probably the reason that Kolodner \cite{kolss2} does not appear to
have seen any evidence for the existence of a critical
$\varepsilon_{\it\!c}^{\it so}$ in his experiments on traveling waves
in binary mixtures, as in this system the transition is weakly
subcritical \cite{bensimon,nonadi}.

\subsection{Sinks}\label{ss_sink}

As we have seen in section \ref{ss_mul}, counting arguments show that there
generically exists a two-parameter family of uniformly translating
sink solutions. The scaling of their width as a function of $\varepsilon$ 
is not completely obvious, since the figures of Cross 
\cite{cross1}\footnote{
  The work of Cross was motivated by experiments on traveling waves in
  binary mixtures. In such systems, the bifurcation is weakly
  subcritical; experimentally, the sinks width is then expected to be
  finite for small $\varepsilon$.} 
indicate that their width approaches
a finite value as $\varepsilon \downarrow 0$, while Coullet {\em et
  al.} found a class of sink solutions whose width diverges as
$\varepsilon^{-1}$ for $\varepsilon \downarrow 0$.  

In appendix \ref{app_epszero} we demonstrate, by examining the ODE's 
(\ref{al}-\ref{zr}) in the $\varepsilon \downarrow 0$ limit, that the asymptotic scaling of the width of sinks as $\varepsilon^{-1}$ follows naturally.

If we now focus again on uniformly translating sink structures of the form
\begin{equation}
A_{R,L}=e^{-i \omega_{R,L} t} {\hat A}_{R,L}(\xi) ~,
\end{equation}
and explicitly carry out this scaling by introducing the scaled variables
\begin{eqnarray}\label{scal}
\bar \xi = \varepsilon \xi~,~~~~~~
\bar \omega_{R,L} = \frac {\omega_{R,L}}{\varepsilon}~,~~~~~~
\bar A_{R,L} = \frac{\hat A_{R,L}}{\sqrt\varepsilon}~,
\end{eqnarray}
We find that, {\em if} the limit $\varepsilon \rightarrow 0$ is regular
we can (to lowest order in $\varepsilon$), approximate the ODE's (\ref{al}-\ref{zr}) by the following reduced set of equations
\begin{eqnarray}
  (-i \bar \omega + s_0 \partial_{\bar \xi})\bar A_R &=& \bar A_R -
  (1-ic_3)|\bar A_R|^2 \bar A_R - g_2(1-ic_2)|\bar
  A_L|^2 \bar A_R\label{scaleq1} \\ 
(-i \bar \omega-s_0 \partial_{\bar \xi}) \bar A_L
  &=& \bar A_L-(1-ic_3)|\bar A_L|^2 \bar A_L -
  g_2(1-ic_2)|\bar A_R|^2 \bar A_L ~,\label{scaleq2}
\end{eqnarray}
where we have set $\bar \omega_R\!=\!\bar\omega_L\!=\!\omega$ and $v\!=\!0$,
to study symmetric, stationary sinks. As one can see by comparing Eqs. (\ref{scaleq1}-\ref{scaleq2}) with the original equations (\ref{al}-\ref{zr}), the taking of the $\varepsilon \rightarrow 0$ limit effectively amounts to the removal
of the diffusive term $\propto \partial_{\xi}^2$. One could {\em a priori} wonder whether this procedure is justified, since we are removing the highest order
derivative from the equations, which could very well constitute a
singular perturbation. This matter will be resolved below with the aid of our counting argument.

Equations (\ref{scaleq1}-\ref{scaleq2}) admit an exact solution for the sink profile, first obtained by Coullet {\em et al}. When we substitute
\begin{equation}
\bar A_{R,L} = \bar a_L e^{i \bar \phi_{R,L}}~,~~~~~~
\bar q_{R,L} = \partial_{\bar \xi} \bar \phi_{R,L}~,
\end{equation}
the explicit solution is given by
\begin{equation}\label{solution}
  a_R (x)= \sqrt{ \frac{\varepsilon} {1+e^{(2(g_2-1)\varepsilon
        x)/s_0}} }= \sqrt{\varepsilon-a_L^2}~.
\end{equation}

The width of these solutions is easily seen to indeed diverge as $\varepsilon^{-1}$. Since we can still vary $\bar \omega$ continuously to give various values for the asymptotic wavenumber, which is for solutions of the type (\ref{solution}) given by
\begin{equation}
   \bar q_R = \frac{1}{s_0} (\bar \omega +c_3) ~ \mbox{for} ~ \bar \xi
   = -\infty ~ \mbox{and }
   \bar q_L = \frac{-1}{s_0} (\bar \omega
   +c_3) ~ \mbox{for} ~ \bar \xi = \infty ~,
\end{equation}
we see that we still have a 1-parameter family of $v\!=\!0$ sinks. Since this is in accord with the full counting argument, the limit $\varepsilon \downarrow 0$ is indeed regular.

In passing we note that source solutions of finite width are
completely absent in the scaled
Eqs. (\ref{scaleq1}-\ref{scaleq2}). This 
is because the only orbit that starts from the $A_R=0$ single mode
fixed point and flows to the $A_L=0$ single mode fixed point passes
through the $A_L=A_R=0$ fixed point, and therefore takes an infinite
pseudo-time $\xi$; such a source has an infinitely wide core regime
where $A_L$ and $A_R$ are both zero.  This also agrees with our
earlier observations, since the coherent sources already diverge at
finite $\varepsilon_{\it\!c}^{\it so}$.

In Fig. \ref{sinkwidth} we plot the sink width versus $\varepsilon$ for the
full set of ODE's, as obtained from our shooting. It is clear that the
sink indeed diverges at $\varepsilon \!=\!0$, and that it
asymptotically approaches the theoretical prediction from the above
analysis. 

\subsection{The limit $s_0 \rightarrow 0$}\label{ss_s0} 

In this paper, we focus mainly on the experimentally most relevant
limit $s_0$ {\em finite}, $\varepsilon $ {\em small}. For
completeness, we also mention that Malomed \cite{malomed} has also
investigated the limit where $\varepsilon$ is nonzero and
$s_0\rightarrow 0$, $c_i \rightarrow 0$, perturbatively. In this
limit, which is relevant for some laser systems \cite{lasers}, sinks are
found to be {\em wider} than sources. This finding can easily be recovered
from the results of our appendix: From (\ref{disp}) it follows that to
first order in $s_0$ the change in the exponential growth rate
$\kappa$ of the suppressed mode away from zero is
\begin{equation}
\label{changekappa}
\delta \kappa_L^\pm = -s_0 /2~,~~~~~\delta \kappa_R^\pm = s_0 /2~.
\end{equation}
where according to our convention of the appendices, $\kappa^-$
corresponds to the 
negative root of (\ref{disp}), and $\kappa^+$ to the positive one.  For a
sink, the left traveling mode is suppressed on the left of the
structure, and so this mode grows as $\exp(\kappa_L^+ \xi)$, while on
the right of the sink the right-traveling mode decays to zero as
$\exp(\kappa_L^- \xi)$. For the sources, the right and left traveling
modes are interchanged.  According to (\ref{changekappa}), upon
increasing $s_0$ the relevant rate of spatial growth and decay
decreases for sinks and increases for sources. Hence in this limit,
somewhat counter-intuitively, sinks are wider than sources. For a
further discussion of the limit $s_0 \rightarrow 0$, we refer to the
paper by Malomed \cite{malomed}.

\section{Dynamical properties of source/sink patterns}\label{s_con}

Apart from the instability of the sources that occurs when
$\varepsilon<\varepsilon_{\it\!c}^{\it so}$, there are at least two
other mechanisms that lead to nontrivial dynamics of source/sink
patterns, and this section is devoted to a description of such states.
Due to the high dimensionality of the parameter space (one has to
consider, in principle, the coefficients $c_1,c_2,c_3,g_2$ and
$\varepsilon$ or $s_0$), we aim at presenting some typical examples
and uncovering general mechanisms, rather than aiming at a complete
overview. Several of the scenario's we lay out deserve further
detailed investigation in the future.

The starting point of our analysis here is the discrete nature of the
sources (see section \ref{ss_mul}) which implies that the wave\-number
of the laminar patches is often uniquely determined
\cite{coullet3,malomed,arandw}. A stability analysis of these
waves yields the two following instability mechanisms:

\begin{itemize}
\item Benjamin-Feir instability. When the waves emitted by the sources
  are unstable to long wavelength modes, it is the nature of this
  instability, i.e., whether it is {\em convective} or {\em absolute},
  that determines the global dynamical behavior.  The dynamical states
  that occur in this case are discussed in section \ref{ss_con}.
\item Bimodal instabilities. The selected wavenumber can also lead to
  an instability resulting from the competition between the left and
  right traveling modes. The essential observation is that for a
  selected wavenumber $q_{sel}$ there exists a range
  $1\!<\!g_2<\varepsilon/(\varepsilon-q_{sel}^2)$ for which {\em both}
  single and bimodal states are unstable. Provided that there are
  sources in the system, we find then a regime of {\em source-induced
    bimodal} chaos (see section \ref{ss_bim}).
\end{itemize}

Furthermore, both of these instabilities can occur simultaneously, as
seems to be the case in experiments of the Saclay group
\cite{daviaud}, and can be combined with the small-$\varepsilon$
instability of the sources, discussed in section \ref{s_sca}.  This
leads to quite a rich palette of dynamical and chaotic states (section
\ref{ss_rich}). We have summarized the various disordered states that
are typical for the coupled amplitude equations in table
\ref{germantable} above. The first three types of dynamics are
source-driven.  Sources are not essential for the last three types of
dynamics, which are driven by the coupling between the $A_L$ and $A_R$
modes.

\begin{table}
\caption{Overview of disordered and chaotic states.}\label{germantable}
\begin{tabular}{llll}
  Type & Section & Fig. & Parameters \\\hline Core-instabilities &
  \ref{ss_san},\ref{ss_num} & \ref{wide} &
  $\varepsilon  < \varepsilon_c^{\it so} = s_0^2/(4+4 c_1^2)$\\
  Absolute instabilities & \ref{ss_con} &\ref{f7},\ref{f8} &$v^*_{BF}>0$\\
  Bimodal chaos & \ref{ss_bim}& \ref{f9}&
  $1<g_2<\varepsilon /(\varepsilon-q_{sel})$\\
  Defects + Bimodal & \ref{sss_pb} &\ref{f10}&
  $g_2$ just above 1\\
  Intermittent + Bimodal & \ref{sss_inter}&\ref{fsti}&$g_2$ just above 1\\
  Periodic patterns & \ref{sss_per}&\ref{f7},\ref{f8},\ref{f11a}&

$c_2$,$c_3$: opposite
  signs and not small\\\hline
\end{tabular}

\end{table}

\subsection{Convective and absolute sideband-instabilities}\label{ss_con}

Plane waves in the single CGL equation with wavenumber $q$ exhibit
sideband instabilities when \cite{ch}\footnote{When both nominator and
  denominator are negative, as may occur for large $c_1$, this
  equation seems to suggest that one might have a stable band of
  wavenumbers. However, when $1-c_1c_3$ is negative, no waves are
  stable; the flipping of the sign of the denominator for large $c_1$
  bears no physical relevance, but is due to a long-wavelength
  expansion performed to obtain Eq. (\ref{nonlq}). Note that the
  denominator is always positive as long as $1-c_1 c_3$ is positive.}
\begin{equation}\label{nonlq}
  q^2 > \frac{\varepsilon(1-c_1c_3)}{3-c_1c_3+2c_3^2}~,
\end{equation}
and when the curve $c_1 c_3\!=\!1$ is crossed, all plane waves become
unstable, and one encounters various types of spatio-temporal chaos
\cite{ch,overview1,overview1b,overview1c}.  For the coupled CGL equations under
consideration here, the condition for linear stability of a single
mode is still given by Eq. (\ref{nonlq}), since the mode which is
suppressed is coupled quadratically to the one which is nonzero.
Since the sources in general select a wavenumber unequal to zero, the
relevant stability boundary for the plane waves in source/sink
patterns typically lies below the $c_1c_3=1$ curve. 

Consider now a linearly unstable plane wave.  Perturbations of this
wave grow, spread and are advected by the group velocity. The
instability of the wave is called convective when the perturbations
are advected away faster than they grow and spread; when monitored at
a fixed position, all perturbations eventually decay. In the case of
absolute instability, the perturbations spread faster than they are
advected; such an instability often results in persistent dynamics. To
distinguish between these two cases one has to compare, therefore, the
group velocity and the spreading velocity of perturbations.  For a
general introduction to the concepts of convective and absolute
instabilities, see e.g. \cite{chomaz_maybe}.

Numerical simulations of the coupled CGL equations presented below
show that the distinction between the two types of instabilities is
important for the dynamical behavior of the source/sink patterns. When
the waves that are selected by the sources are convectively unstable,
we find that, after transients have died out, the pattern typically
``freezes'' in an irregular juxtaposition of stationary sources and
sinks.  When the waves are absolutely unstable\footnote{It should be
  noted that the criterion for absolute instability concerns the
  propagation of perturbations in an ideal, homogeneous background.
  For typical source/sink patterns, one has finite patches; the
  criterion can also not determine when perturbations are strong
  enough to really affect the core of the sources.  Analogous to the
  2D case, we have found that persistent dynamics sets in slightly
  {\em above\/} the threshold between convective and absolute
  instabilities.}, however, persistent chaos occurs.

The wavenumber selection and instability scenario sketched above for
the coupled CGL equations is essentially the one-dimensional analogue
to the ``vortex-glass'' and defect chaos states in the 2D CGL equation
\cite{spiralrefs,spirabscon}; in that case the wavenumber is selected
by so-called spiral or vortex solutions.  As we shall
discuss, there are, however, also 
some differences between these cases.

Below we will briefly indicate how the threshold between absolute and
convective instabilities is calculated (see also \cite{spirabscon}).
The advection of a small perturbation is given by the nonlinear group
velocity $s=\partial \omega /\partial q$ which is the sum of the
linear group velocity $s_0$ and the nonlinear term $s_q\!:=\!2 q (c_1
+ c_3)$:
\begin{equation}
s_L = - s_0 + 2 q_L (c_1 + c_3)~,~~
s_R =   s_0 + 2 q_R (c_1 + c_3)~.\label{vg2}
\end{equation}
The spreading velocity of perturbations is conveniently calculated in
the linear marginal stability/pulled front
 framework \cite{lmsc,ebert}  once one has
obtained a dispersion relation for these perturbations.  Since we
consider single mode patches, we are allowed to restrict ourselves to
a single CGL equation, in which the linear group velocity term $\pm
s_o \partial_x A$ is easily incorporated, as it just gives a constant
boost.  Considering a perturbed plane wave of the form $A\!=\! (a+ u)
\exp{i(qx-\omega t)}$, where $u$ is a small complex-valued
perturbation $\sim \exp{i(k x -\sigma t)}$ and $a^2 \!=\!
\varepsilon- q^2$.  Upon substituting this Ansatz into a single CGL
equation, linearizing and going to a Fourier representation, one
obtains a dispersion relation $\sigma(k)$ \cite{TomasBook}. From this
relation one then finally calculates the spreading velocity $v^*_{BF}$ of
the Benjamin-Feir perturbations in the linear marginal stability or
saddle-point framework \cite{lmsc}.

Since in general we can only calculate the selected wavenumber $q$ by
a shooting procedure of the ODE's (\ref{al}-\ref{zr}) for a source,
obtaining a full overview of the stability of the plane waves as a
function of the coefficients necessarily involves extensive numerical
calculations. Therefore, we will focus now on a single sweep of $c_2$.
For reasons to be made clear below, we choose
$\varepsilon\!=\!1,c_1\!=\!c_3\!=\!0.9, s_0\!=\!0.1$ and $g_2\!=\!2$.
Since we fix all coefficients but $c_2$, the stability boundary
(\ref{nonlq}) is fixed.  By sweeping $c_2$, the selected wavenumber
varies over a range of order 1, and one encounters both convective and
absolute instabilities.

We have found that after a transient, patterns in the stable or
convectively unstable case are indistinguishable\footnote{Except, of
  course, when we prepare a very large system with widely separated
  sources and sinks}. When there is no inherent source of noise or
perturbations, there is nothing that can be amplified, and the
convective instability is rendered powerless (see however, section
\ref{ss_rich}).

Although the transition between stable and convectively unstable waves
is not very relevant for the source/sinks patterns here, the
transition between convectively and absolutely unstable waves is
interesting.  To obtain an absolute instability one needs to carefully
choose the parameters; when $q$ increases, the contribution to the
group velocity of the nonlinear term $s_q$ increases, and we have to
take $c_1$ and $c_3$ quite close to the $c_1 c_3 \!=\!1$ curve to find
absolute instabilities.  This is the reason for our choice of
coefficients.  In Fig. \ref{f5} we have plotted the selected frequency
(obtained by shooting), corresponding wavenumber and propagation
velocity $v_{BF}^*$ of the mode to the right of the source, as a
function of $c_2$.  For this choice of coefficients the single mode
waves turn Benjamin-Feir convectively unstable when, accordingly to
Eq. (\ref{nonlq}) $|q| > 0.223$ , which is the case for all values of
$c_2$ shown in Fig. \ref{f5}.  The waves turn absolutely unstable when
$|q| >0.553$, and this yields that the waves become absolutely
unstable for $c_2<-0.25$.

When the selected waves becomes absolutely unstable, the sources may
be destroyed since perturbations can no longer be advected away from
them; the system typically ends up in a chaotic state.  In Fig.
\ref{f7} we show what happens when we choose the coefficients as in
Fig. \ref{f5}, and decrease $c_2$ deeper and deeper into the
absolutely unstable regime.  All runs start from random initial
conditions, and a transient of $t=10^4$ was deleted.  Although the
left- and right traveling waves do not totally suppress each other, it
was found that pictures of $|A_L|$ and $|A_R|$ are, to within good
approximation, each others negative (see also the final states in Fig.
\ref{f8}).  In accordance with this, we choose our greyscale coding to
correspond to $|A_R|$, such that light areas corresponds to
right-traveling waves and dark ones to left-traveling waves.

In Fig. \ref{f7}a, $c_2\!=\!-0.3$ and the waves have just turned
absolutely  unstable, but the only nontrivial dynamics is a very slow
drift of some of the sources and sinks. Note that this does not
invalidate our counting results that isolated sources are typically
stationary, because the drifting occurs only for structures that are
close together. When $c_2$ is lowered to $-0.4$ (Fig. \ref{f7}b), one
can see now the Benjamin-Feir perturbations spreading out in the
opposite direction of the group velocity, eventually affecting the
sources (for example around $x\!=\!230, t\!=\! 2700$). Some of the sinks 
become very irregular. When $c_2$ is decreased even further to $-0.6$
(Fig. \ref{f7}c), the sources and sinks show a tendency to form
periodic states \cite{saka1} (see also Fig. \ref{f8}). These states seem
 at most weakly unstable since only some very mild oscillations
are observed. The two sinks with the largest patches around them
show most dynamics, and one sees the irregular creation and annihilation
of small source/sink pairs here (around $x\!=\!320$ and $440$).
Finally, when $c_2$ is decreased to $-0.8$ (Fig. \ref{f7}d) the state
becomes more and more disordered; the irregular ``jumping'' sink at
$x\approx 230$ is worth noting here.

It is interesting to note that, in particular for large negative
$c_2$ closely bound, uniformly drifting sink-source pairs are formed
(see for instance around $x\!=\!430,t\!=\!700$ in Fig. \ref{f7}d).
Another frequently occurring type of solution are periodic states,
corresponding to an array of alternating patches of $A_L$ and $A_R$
mode (see also Fig. \ref{f8}).  The source/sink pairs and in
particular the periodic states occur over a quite wide range of
coefficients; their existence has been reported before by Sakaguchi
\cite{saka1}.  In a coherent structures framework, periodic states
correspond to limit cycles of the ODE's (\ref{al}-\ref{zr}).  In many
cases they can be seen as strongly nonlinear standing waves, and they
show an interesting destabilization route to chaos (see section
\ref{sss_per}).

Apart from the similarities between the mechanisms here and the spiral
chaos of the 2D CGL equation, it is also enlightening to notice the
differences. The first difference is that our sources, in contrast to
the spirals in 2D, are not topologically stable. In the states we have
shown so far this does not play a role; in the following section we
will see examples where instabilities of the sources themselves play a
role.  While in the 2D case the spiral cores that play the role of a
source are created and annihilated in pairs, it is here only the
sources and sinks that are created or annihilated in pairs.
Furthermore, in the spiral case, the linear analysis that determines
whether the waves are absolutely of convectively unstable is performed
for plane waves. This means one neglects curvature corrections of the
order $1/r$, where $r$ is the distance to the core of the
source. Here, the only correction comes from the asymptotic,
exponential approach of the wave to a plane wave; this exponential
decay rate is given by the decay rate $\kappa$ (see the
appendix).  Finally, in the spiral case, for fixed $c_1$ and $c_3$,
both the group velocity and the selected wavenumber are fixed, while
here the selected wavenumber can be tuned by $c_2$, without
influencing the stability boundaries of the single mode state. The
group velocity can be tuned by $s_0$. Although the selected wavenumber
influences the group velocity, cf. Eqs.  (\ref{vg2}), and
$s_0$ influences the selected wavenumber, this large number of
coefficients gives us more freedom to tune the instabilities.

\subsection{Instability to bimodal states: 
source-induced bimodal chaos}\label{ss_bim}

The dynamics we study in this section is intrinsically due to a
competition between the single source-selected waves and bimodal
states. Therefore, this state is in an essential way different from
what can be found in a single CGL equation framework.

The wavenumber selection by the sources is of importance to understand
the competition between single mode and bimodal states.  In the usual
stability analysis of the single mode and bimodal states, it is
assumed that both the $A_L$ and $A_R$ modes have equal wavenumber
\cite{coullet}. Therefore, this analysis does not apply to the case of
a single mode, say the right-traveling mode, with nonzero wavenumber.
The left-traveling mode is then in the zero amplitude state and has no
well-defined wavenumber; one should consider therefore its fastest
growing mode, i.e., a wavenumber of zero.  The following, limited
analysis, already shows that for $g_2$ just above 1, instabilities are
expected to occur.  Restricting ourselves to long wavelength
instabilities, the analysis is simply as follows.  Write the left- and
right-traveling waves as the product of a time dependent amplitude and
a plane wave solution:
\begin{equation}
A_L= a_L(t) e^{i(q_L x - \omega_L t)}~,~~~
A_R= a_R(t) e^{i(q_R x - \omega_R t)}~,
\end{equation}
and substitute this Ansatz in
the coupled CGL equations. One obtains then  the following set of ODE's
\begin{equation}
\partial_t a_L = (\varepsilon - q_L^2 - a_L^2 - g_2 a_R^2) a_L~,~~~~
\partial_t a_R = (\varepsilon - q_R^2 - a_R^2 - g_2 a_L^2) a_R~.\label{gt}
\end{equation}
Consider the single mode state with $a_R\!\neq\!0, a_L\!=\!0$ and take
$q_L\!=\!0$. The maximum linear growth rate of $a_L$ now follows from
Eq. (\ref{gt}) to be the one with $q_L\!=\!0$; this mode has a
growth rate given by $\varepsilon -g_2 a_R^2 \!=\!  \varepsilon-g_2
(\varepsilon - q_R^2)$. From this it follows that a single mode state
with wavenumber $q_R$ is unstable when $g_2<\varepsilon/(\varepsilon-
q_R^2)$.  In source/sink patterns, the selected wavenumber is as large
as $\sqrt{\varepsilon/3}$ at the edge of the stability band for
$c_1\!=\!c_3\!=\!0$; it is as large as $0.6\sqrt{\varepsilon}$ in Fig.
\ref{f5}.  In extreme cases, the value of $g_2$ necessary to stabilize
plane waves can be at least 50\% larger than the value $1$ that one
would expect naively.

On the other hand, the stability analysis of the bimodal states shows
that they are certainly unstable for $g_2\!>\!1$.  A na\"{\i}ve analysis
for general $q_L$ and $q_R$, based on Eqs. (\ref{gt}) can be performed
as follows.  Solving the fixed point equations of Eqs.  (\ref{gt}) for
the bimodal state (i.e., $a_L$ and $a_R$ both unequal to zero), and
linearizing around this fixed point yields a $2\times2$ matrix.  From
an inspection of the eigenvalues we find that the bimodal states turn
unstable when $g_2<\varepsilon-q_1^2/(\varepsilon-q_2^2)$, where $q_1$
is the largest and $q_2$ is the smallest of the wavenumbers $q_L,
q_R$.  When both wavenumbers are equal this critical value of $g_2$ is
one; it is smaller in general.  

It should be noted that this analysis does not capture sideband
instabilities that may occur, and therefore waves in a much wider
range might be unstable. For sideband-instabilities of bimodal states,
the reader may consult \cite{coullet} and \cite{biriecke}.  However,
our analysis shows already that there is certainly a regime around $g_2=1$
where {\em both} the single and bimodal states are unstable.  This
regime at least includes the range $1<g_2<\varepsilon/(\varepsilon-
q_{\it sel}^2)$.

The distinction between convective and absolute instabilities becomes
slightly blurred here. Suppose for instance we inspect a single-mode
state that turns unstable against bimodal perturbations.  Initially,
these perturbation will be advected by the group velocity of the
nonlinear mode, but as the perturbations grow, both modes will start
to play a role, and since they feel a group velocity of opposite sign,
the perturbations are effectively slowed down. Roughly speaking, the
instability might be linearly convectively unstable but nonlinearly
absolutely unstable \cite{chomaz_maybe}.

Without going into further detail we will now show two examples of the
bimodal chaos that occurs when $g_2$ is just above 1.  For examples of
similar dynamics, also for $g_2<1$, see \cite{biriecke}.  In the first
example (Fig. \ref{f9}a-b) we have taken
$\varepsilon\!=\!1,c_1\!=\!c_3\!=\!0.5,c_2\!=\!-0.7,s_0\!=\!1$ and
$g_2\!=\!1.1$. The selected wavenumber is almost independent of the
value of $g_2$ and approximately equal to $0.35$, which yields a
critical value of $g_2$ of 1.14. For $g_2$ just below this value, the
instability appears convective, and after a transient the system ends
up in a mildly fluctuating source/sink pattern. When $g_2$ is
decreased, the instability becomes stronger and, presumably, absolute
in nature.  The {\em sources} behave then very irregularly, while the
sinks drift according to there incoming, disordered waves. Note that
sources and sinks are created and annihilated in this state.  In Fig.
\ref{f9}c-d we show the disordered dynamics for $
\varepsilon\!=\!1,c_1\!=\!1,c_3\!=\!-1,c_2\!=\!1,s_0\!=\!0.5$ and
$g_2\!=\!1.1$.  Note that in the laminar patches, since
$c_1\!=\!-c_3$, the dynamics is relaxational \cite{ch,vhhvs}.  In this
state, no creation or annihilation of sources and sinks is found; the
sinks drift slowly, while the sources behave very irregularly.

The dynamical states as shown in Fig. \ref{f9} are different from the
chaotic states that we are familiar with from the single CGL equation
\cite{overview1,overview1b,overview1c,homoclons}, and so they are of
some interest in their own right.  Note that it is possible to get
persistent dynamics for values of $c_1$ and $c_3$ that in a single CGL
equation-framework would lead to completely orderly dynamics.  As the
two examples in Fig. \ref{f9} show, qualitatively different states
seem to be possible in this regime; the question of classification of
the various dynamical states is completely open as far as we are
aware.

Finally, it should be pointed out that when, as is the case here, the
left- and right-traveling mode no longer suppress each other,
$\varepsilon_{\it\!eff\!}$ becomes positive.  In principle this might
change the multiplicity of the sources, since the eigenvalues coming
from the linear fixed point can have a different structure for
positive $\varepsilon_{\it\!eff\!}$ (see appendix \ref{eso_eeff}).
However, this is only true when the effective velocity $v\pm s_0$ is
larger than the critical velocity $v_{cL}$; for the cases considered
above, this does not happen. Hence, the sources are here still unique
and select a unique wavenumber.

\subsection{Mixed mechanisms}\label{ss_rich}

In the previous sections we have described three mechanisms by which
sink/\-source patterns can be destabilized.  First of all, in section
\ref{s_sca} we found that due to a competition between the linear
group velocity $s_0$ and the propagation of linear fronts, the
cores of the sources become unstable when
$\varepsilon\!<\!\varepsilon_c^{so}$. 
In section \ref{ss_con} we have shown that the waves that are
sent out by the sources can be convectively or even absolutely
unstable, and in section \ref{ss_bim}
we found that these waves may also be unstable to bimodal
perturbations when $g_2$ is not very far above $1$.
Since the mechanisms that lead to these instabilities
are independent,  these instabilities might occur together. This is
the subject of this section.
In particular, one can always lower the control parameter
$\varepsilon$ in an experiment to make the sources become
core-unstable (section \ref{sss_ciua}).  A second combination of
instabilities occurs when $g_2$ is close to 1 and the plane waves are
unstable and generate phase slips (section \ref{sss_pb}); a particular
interesting case occurs when the single mode waves are in the
so-called intermittent regime (section \ref{sss_inter}).

\subsubsection{Core instabilities and unstable waves}\label{sss_ciua}
As discussed in section \ref{ss_num}, the cores of the source may
start to fluctuate when $\varepsilon\!<\!\varepsilon_c^{so}$.  As is
visible in Fig. \ref{wide}c, the perturbations that are generated in
the core are then advected away into the asymptotic plane waves.  In
the discussions in section \ref{s_sca} above, we have focused on the
case where these waves are stable, but obviously, when they are
unstable, this will amplify the perturbations emitted by the source
core. In particular, when the waves are convectively unstable, a
stable core for $\varepsilon\!>\!\varepsilon_c^{so}$ leads to
stationary patterns, but a fluctuating core can fuel the convective
instabilities. This yields a simple experimental protocol to check for
convective instabilities; simply lower $\varepsilon$ and follow the
perturbations send by the sources for
$\varepsilon\!>\!\varepsilon_c^{so}$.

\subsubsection{Phase slips and bimodal instabilities}\label{sss_pb}

Let us for definiteness suppose we have that $A_L\!=\!0$, and the
right-traveling mode is active.  When this $A_R$ mode is chaotic and
displays phase slips, the effective growth rate of the $A_L$ mode,
$\varepsilon^L_{\it eff}$, may become positive for some period.  $A_L$
only grows during this period; it depends then on the duration and
spatial extension of the positive $\varepsilon^L_{\it eff}$ ``pocket''
whether $A_L$ can grow on average. Clearly, one should look at a
properly averaged value of $\varepsilon^L_{\it eff}$, and therefore at
the averages of $\varepsilon - g_2 a_R^2$ \cite{saka1}.  When $g_2$ is
sufficiently large, the averaged effective growth rate always becomes
negative, so that even a heavily phase slipping wave can still
suppress its counter-propagating partner.

We show two examples of the dynamics when phase slips occur and $g_2$
is not large enough to strictly suppress the near-zero mode. As
coefficients we choose
$c_1\!=\!1,c_3\!=\!1.4,c_2\!=\!1,\varepsilon\!=\!1, s_0\!=\!0.5$, and
the dynamics is illustrated in Fig. \ref{f10}.

It should be noted that in Fig. \ref{f10}b the sources are stationary,
while some of the sinks drift. This seems to be due to the fact
that near the sink, i.e., far away from the sources, the wave emitted
by the sources has undergone phase slips, and the incoming wavenumbers
of the sink can therefore be different from the source-selected
wavenumbers. For slightly different coefficients we have observed
patterns of stationary sources, with sinks in between that by this
mechanism move in zig-zag fashion, i.e., alternating to the left and
to the right.

\subsubsection{Intermittency and bimodal instabilities}\label{sss_inter}

Recently, Amengual {\em et al.} studied the case of spatio-temporal
intermittency in the coupled CGL equations for a linear group velocity
$s_0\!=\!0$ and $c_2\!=\!c_3$ \cite{amen}.  This particular sub-case
of the coupled CGL equations is of importance in the description of
some laser systems \cite{amen,lasers}.  When $g_2$ is increased from
zero, the authors of \cite{amen} found that for $g_2<1$ one finds
intermittency, with the $A_L$ and $A_R$ obviously becoming more and
more correlated as the cross-coupling increases.  Furthermore, the
authors observed that for $g_2>1$, the two modes become
``synchronized'', i.e., the intermittency disappears and the systems
ends up in a state that we recognize now as a stationary source/source
pattern (not source/sink, see below).  Since the intermittency
``disappears'' the authors question the applicability of a single CGL
equation for patches of single modes in the coupled CGL equations
(\ref{coupcgl1}-\ref{coupcgl2}).

The purpose of this section is to clarify, correct and extend their
results, using our results for the wavenumber selection, the bimodal
instabilities and the discussion in section \ref{sss_pb}. In
particular we will show that, {\em(i)} for sufficiently large $g_2$,
the intermittency can persist, {\em(ii)} when the intermittency
disappears it can do so by at least two distinct mechanisms,
{\em(iii)} more complicated states can occur.  We conclude then that
for single mode patches the single CGL is a correct description,
provided one is sufficiently far away from bimodal instabilities and
one takes the source-selected wavenumber and correct boundary
conditions into account.

For the case considered in \cite{amen} the group-velocity $s_0$ is
equal to zero, so the two modes $A_L$ and $A_R$ are completely
equivalent. The distinction between sources and sinks depends
therefore on the nonlinear group velocity, which follows from the
selected wavenumber. The counting arguments yield in this case again a
discrete $v\!=\!0$ source and a two parameter family of sinks (see
section \ref{ss_coh}).  In simulations we typically find stationary
sources that separate the patches of $A_L$ and $A_R$ mode, and {\em
  single amplitude sinks} sandwiched in between these sources.

We will show now a variety of scenarios for intermittency in the
coupled CGL equations (\ref{coupcgl1}-\ref{coupcgl2}).  The
coefficients used in \cite{amen} are $c_1\!=\!0.2, c_2\!=\!c_3\!=\!2,
\varepsilon\!=\!1$ and $s_0\!=\!0$.  The coefficients $c_1$ and $c_3$
are chosen such that a single mode is in the so-called intermittent
regime. In this regime, depending on initial conditions, one may
either obtain a plane wave attractor or a chaotic, ``intermittent''
state; the latter one is typically built up from propagating
homoclinic holes and phase slips
\cite{overview1,overview1b,overview1c,homoclons}. 

In Fig. \ref{fsti}a we take $g_2=2$ and start from an ordered pair of
sources. By a rapidly changing $c_1$ to a value of $1.2$ and then back
to the value $0.2$, we generate phase slips that nucleate a typical
intermittent state. This intermittent state persists for long times;
there is no ``synchronization'' whatsoever.  We found that we can also
first let the source develop completely, and then introduce some phase
slips; also in this case the intermittency clearly persists.  To
understand this, note that in this case $g_2$ is sufficiently large,
and so $\varepsilon_{\it\!eff\!}$ is negative (see \ref{sss_pb});
although there are phase slips, the two modes suppress each other
completely.

In contrast, when $g_2$ is lowered, $\varepsilon_{\it\!eff\!}$ can
become positive, and this corresponds to the scenario described in
\cite{amen}. In Fig. \ref{fsti}b we start from state obtained for
$g_2$=2, and then quench $g_2$ to a value of $1.5$. In this case,
$\varepsilon_{\it eff}$ becomes positive every now and then, and after
a while, in the patch originally the exclusive domain of $A_L$, small
blobs of $A_R$ mode grow. After a sufficient period has elapsed, these
blobs nucleate new sources, and the system ends up in a stationary
source/source pattern.  The laminar patches in between the sources are
quite small and the intermittency disappears.  

The system switches from the intermittent to the plane wave attractor
when the new sources are formed; this does not mean that the CGL
equation is incorrect here, since both plane waves and intermittent
states are attractors for these coefficients.  The dissappearance of
the intermittency can be easily understood as follows: the main
mechanism by which intermittency spreads through the single CGL
equation is by the propagation of homoclinic holes that are connected
by phase slip events \cite{homoclons}. If the phase slips now generate
sources, there is no generation of new homoclinic holes and the
intermittency dies out.  

It should be noted that for this particular choice of the coefficients
$c_1$ and $c_3$, the homoclinic holes have a quite deep minimum in
$|A|$, which increases the value of the average of $\varepsilon_{\it
  eff}$; therefore one needs quite a large $g_2$ to guarantee the
mutual suppression of the $A_L$ and $A_R$ modes.

Finally, we found that the selected wavenumber for the coefficients of
this particular example is $\approx 0.1$.  As a consequence, the
transition to stationary domains as observed in \cite{amen} can {\em
  not} occur at $g_2$ precisely equal to $1$, but occurs for $g_2
\approx 1.01$ (see section \ref{ss_bim}).

This generation of sources due to phase slips of the nonlinear mode is
not the only way in which the intermittency can disappear.  Considerer
the example shown in Fig. \ref{fsti}c. We have chosen the coefficients
as $c_1\!=\!0.6,c_3\!=\!1.4,c_2\!=\!1,\varepsilon\!=\!1,s_0\!=\!0.1$
and $g_2\!=\!2$. The sources select now a wavenumber of $ 0.3783$, and
the plane wave emitted by the source simply ``eats up'' the
intermittent state; note the single amplitude sinks visible for late
times.  It should be realized that many dynamical states are sensitive
to a background wavenumber, and that the spatio-temporal intermittent
state is particularly sensitive to this \cite{homoclons}; when
describing a patch in the coupled CGL equations by a single CGL
equation, one should take into account that one has wave-selection at
the boundaries due to the sources.

Finally, when $c_2$ is lowered to a value of $0$, the sources
themselves become unstable and the system displays the tendency to
form periodic patterns; these are however not stable, and an example
of the peculiar dynamical states one finds is shown in Fig.
\ref{fsti}d.

In conclusion, when one is far away from any bimodal instabilities,
i.e., when $g_2$ is sufficiently large, a description in terms of a
single CGL equation is sufficient for the patches separating the
sources, provided one takes into account the group velocity, boundary
effects and, most importantly, the selected wavenumber.  It is amusing
to note that the question under which conditions a single amplitude
equation is a correct description of these waves depends on the
coefficients $g_2$ and $c_2$ of the {\em cross-coupling} term.


\subsubsection{Periodic and other states}\label{sss_per}

We would like to conclude this section by showing an example of the
wide range of different states that occur in the coupled amplitude
equations when we sweep $c_2$. We choose the other coefficients as
follows: $g_2\!=\!1.1,c_1\!=\!0.9,c_3
\!=\!2,s_0\!=\!-0.1,\varepsilon\!=\!1$.  Our main finding is that for
large positive or negative $c_2$, their is no sustained dynamics,
while for small $c_2$ we find a strongly chaotic state. In between
there are at least two transitions between laminar and disordered
state (see Figs. \ref{f11a} and \ref{f11b}).

For sufficiently negative $c_2$, all initial conditions evolve to a
spatially periodic state, with rapidly alternating $A_L$ and $A_R$
patches.  We can view these states as an example of highly nonlinear
standing wave patterns.  Depending on initial conditions, these states
may either be stationary or have a small drift. For our particular
choice of coefficients it is empirically found that these states are
linearly stable for $c_2 \leq -0.72$.  In Fig. \ref{f11a}a we see the
evolution from a slightly perturbed initial condition for this value
of $c_2$.  Qualitatively, we observe that when the ``local
wavenumber'' of the standing wave is lowered, this leads to
oscillations, that may or may not lead to ``defects''. After some
reasonably long transient (note the perturbation at $x\approx320,
t\approx 2600$), the dynamics settles down in a slowly drifting
standing wave. This shows that these generalized standing waves are
stable here.

In Fig. \ref{f11a}b we start from such a coherent standing wave state
and have lowered $c_2$ to a value of $-0.71$. In this case
perturbations of the waves are spontaneously formed, indicating a
linear instability. Since the state is unstable, these perturbations
then spread to the system in a way that is reminiscent of the
intermittent patterns obtained, for instance, in experiments on
intermittency in Rayleigh-B\'enard convection \cite{STI2rb}. It should be
noted that, due to the instability of the laminar state, one does not
have an absorbing state, so strictly speaking this state should not be
referred to as intermittent.  Interestingly enough, the
transition between laminar and chaotic behavior seems to be second
order, i.e., we could not find any hysteresis. The transition is
simply triggered by the linear stability of the periodic/standing
waves, and when these waves are stable, they are the only type of
attractor.
 
If $c_2$ is further increased to a value of $-0.5$ (Fig. \ref{f11a}c),
we find a state that we might call defect-chaos of a standing wave
pattern. For $c_2\!=\!0$ (Fig. \ref{f11a}d), the dynamics evolves on
much faster time-scales, and no clear structures are visible by eye.

On the other hand, when we keep increasing $c_2$, we again find regular
states, but these ones correspond to stationary source/sink patterns.
This is illustrated in Fig. \ref{f11b}, where we show four
space-time plots for increasing, positive values of $c_2$.  In
comparison with the dynamics as shown in Fig. \ref{f11a}d, the time
scales become slower and slower when $c_2$ is increased.  This slowing
down becomes quite clear for $c_2\!=\!0.8$ (Fig. \ref{f11b}a) and
$c_2\!=\!0.9$ (Fig. \ref{f11b}b). For $c_2=0.95$ (Fig.  \ref{f11b}c),
the dynamics becomes even more slow and regular. We clearly see now
stationary sources, with irregularly moving sinks in between. Due to
the smallness of $g_2$, phase slips in one of the single modes leads
in some case to the formation of new sources and sinks.
 When $c_2$ is increased to
a value of $1$ (Fig. \ref{f11b}d), some slow dynamics sets in, that
may or may not be a long living transient.
For values of $c_2$ above $1.1$, all initial conditions seem to evolve
to a stationary, regular source/sink pattern.

\section{Outlook and open problems}\label{s_dis}

In this paper we have extended the coherent structures framework and
the counting arguments to the coupled CGL equations, and obtained
important information on the dynamical states that are independent of
the precise values of the coefficients and bear experimental
relevance. In general, these considerations lead to the conclusion
that sources are often unique, sometimes come in pairs but in any case
are at most members of a discrete set of solutions. As a result, they
are instrumental for the wavenumber selection of both regular and
chaotic patterns. Many of the instability mechanisms and dynamical
regimes of the coupled CGL equations can be understood qualitatively
from this point of view, and we have shown several examples of
hitherto unexplored regimes of persistent spatio-temporal chaotic
dynamics (see Table. 1). In this closing section, we wish to discuss
some of these findings in the light of experimental observations, and
summarize the most important open theoretical problems.

\subsection{Experimental Implications}

In short, the experimental predictions that we make, based on our study of the
coupled CGL equations are the following :

$\bullet$ {\em Multiplicity.} Our analysis shows that sources are
expected to come in a discrete set, which would experimentally amount
to a {\em unique}, stationary source.  Furthermore, this source is
expected to be {\em symmetric}, in that it sends out waves of the same
wavenumber to both sides.

Sinks are non-unique. This means that one could have sinks with different
velocities present at the same time. In light of the previous remark on
the uniqueness of sources, this might prove hard to observe experimentally.

$\bullet$ {\em Wavenumber selection.}
One important consequence of the uniqueness of sources is that they select
an asymptotic wavenumber, just as spirals do in the 2D-case. Since the
traveling-wave system is quasi-one-dimensional however, we expect the
wavenumber selection to be much easier to study.

$\bullet$ {\em Scaling Behavior.} We have made definite predictions for
the small-$\varepsilon$ scaling of the width of sources and sinks.
Moreover, we predict the stationary sources to disappear at some
finite value of $\varepsilon$, which is the point where the
non-stationary sources take over. These sources scale as
$\varepsilon^{-1}$, as do the sinks.

$\bullet$ {\em Instabilities and Dynamical Behavior.}
Apart from the non-stationary sources that occur when $\varepsilon$ is
decreased sufficiently, we have found that there are at least two
other mechanisms leading to dynamical states.  The central observation
is that the waves that are selected and sent out by the sources
may become unstable. 
Similar to what happens in the single CGL equation, these waves
can become convectively or absolutely unstable; the latter case in
particular 
yields chaotic states (section \ref{ss_con}). 
When the cross-coupling coefficient is not too far above
one, and the selected wavenumber is unequal to zero, there is
a regime where both single and bimodal states are unstable.

\subsection{Comparison of results with experimental data} \label{ss_com}

Most research in the field of traveling wave systems has
focussed on the properties of the single-mode states, i.e., the states
where the entire experimental cell is filled up by either the left- or
the right-traveling wave. From such a perspective, it is natural to
disregard the source/sink patterns that generally occur initially
above onset as unwanted transient states. Consequently they have not been
studied as extensively as we think they deserve to be.  It is the aim
of this section to confront a number of the theoretical findings of
this article with some of the experimental observations in the heated
wire experiments \cite{alvarez,dubois1,dubois2,dubois3} and in the
experiments on traveling waves in binary liquids
\cite{kolss,kolss2,kolss3,moses,binliq1,binliq2}.  In no way do we claim this
comparison 
to be exhaustive --- the main aim of our discussion is to show that
our results put various earlier observations in a new light, and that
it should be feasible to settle various of the issues we raise with
further systematic experiments.
   
\subsubsection{Heated Wire Experiments}\label{sss_hw}

When a wire which is put a distance of the order of a millimeter under
the free surface of a liquid layer is heated, traveling waves occur
beyond some critical value of the heating power
\cite{alvarez,dubois1,dubois2,dubois3}.  This bifurcation towards
traveling waves turns out to be supercritical \cite{dubois3}, and the
group velocity and phase velocity turn out to have the same sign in
the experiments \cite{alvarez}\footnote{Fig. 11 of \cite{dubois3} also
  illustrates quite nicely that the group velocity and phase velocity
  are parallel.}. The paper by Vince and Dubois \cite{dubois3} is one
of the few papers we know of that discusses the
$\varepsilon$-dependence of the width of sources. The authors show
that the inverse width scales linearly with the heating power $Q$, and
associate this with a scaling of the source width as
$\varepsilon^{-1}$. This is correct if the value of $Q$ at which the
source width diverges coincides with the threshold value for the
linear instability, but whether this is actually the case is
unfortunately not quite clear from the published data\footnote{ In
  the experiments shown in Fig. 10 of \cite{dubois3}, the source width
  diverged at $Q\approx 4.2$ Watts.  Unfortunately, the distance $h$
  between the wire and the fluid surface is not given for the data
  shown. All other measurements in the paper are made at $h=1.34$ mm
  and $h=1.97$ mm, and these values correspond to $Q_c \approx 2.5 $
  Watts and $Q_c \approx 2 $ Watts.} .  Formulated differently, in
terms of our numerical data shown in Fig.  \ref{wide}d, the question
arises whether in the experiments the approximate linear scaling of
the inverse width with the heating power was associated with that of
the thick line above $\varepsilon_c^{so}$, or with the linear scaling
$\sim \varepsilon$ below $\varepsilon_c^{so}$.  If indeed the
experiments are consistent with an $\varepsilon^{-1}$ scaling of the width, then
according to our analysis the sources should be (weakly)
non-stationary and prone to pinning to inhomogeneities in the cell. If
the source width diverges at a finite value of $\varepsilon$, this
might be evidence for the existence of the critical value
$\varepsilon^{so}_c$. It should be of interest to investigate this
further.

In \cite{dubois1}, Dubois {\em et al.} also note that ``$\ldots${\em
  sources may be large when the sinks are always very narrow
  $\ldots$}'' in their heated wire experiments. This agrees with our
finding that sinks are always less wide than the sources but the
published data do not allow us to extract the scaling of the sink width
with $\varepsilon$.

In the experiments by Alvarez {\em et al.} \cite{alvarez}, sources
were found to be stationary and symmetric but non-unique, i.e., each
source sends out the same waves to both sides, but different sources
send out different waves. As a result, patches with different
wavenumbers were found to be present in the system (at any one time),
and the sources were seen to move in response to the fact that they
were sandwiched between waves of different frequency. We have already
seen in section \ref{eso_mult} that there are certain regions of parameter
space where there were two different sources present at the same time
(for one of them, the linear group velocity $s_0$ and nonlinear group
velocity $s$ had opposite signs). However, the fact that we can have
various discrete source solutions can not explain the experimental
observations. First of all, in our simulations two of such sources
were separated by a sink-type structure in one single mode patch, {\em
  not} by a sink separating two oppositely traveling waves, as in the
experiments. Secondly, in the experiments there were always slight
differences between any two pair of sources, which appears
inconsistent with the existence of a finite number of discrete source
solutions.

It appears likely to us that the occurrence of slight differences
between different sources results from the fact that there are always
some impurities or inhomogeneities present in any experimental setup.
Very much like the spirals and target patterns one encounters in the
2D CGL equation \cite{targets}, coherent structures might well be
pinned to such imperfections\footnote{An example of how sources can
  be pinned near cell boundaries below $\varepsilon^{so}_c$ is
  discussed in \cite{rovin}.}. This would of course not invalidate the
results of the counting arguments for the homogeneous case, as it is
precisely on the basis of this counting argument that one would expect
the properties of the discrete source solution(s) to depend
sensitively on the local parameter values.

The sinks which in the experiments of \cite{alvarez} were sandwiched
between two patches with different wavenumbers, were found to move
according to what was referred to as a ``phase matching rule'': during
the motion, a constant phase difference is maintained across the sink
profile, so that no phase slip events occur. This commonly occurs for
sinks in the {\em single} CGL equation, and Fig. \ref{fdouble} 
provides an example of this, but there is one important
difference here: sinks in the experiments separate two oppositely
traveling waves, so phase matching in the actual experiments involves
the {\em fast} scales represented by the critical wavelength $q_{\it
  c}$ of the pattern at onset. In the amplitude approach all
information about this $q_{\it c}$ is lost since we eliminated the
fast scales and only consider the difference between the actual
wavenumber $q$ of the pattern and this $q_c$. At least in the
experiments of \cite{alvarez} the coupling between the fast and the
slow scales is important. These so-called {\em non-adiabatic} effects
\cite{nonadi} will be the object of further study. Experimentally,
it is not clear whether the ``phase matching rule'' was a peculiarity
of \cite{alvarez}, or whether it holds quite generally.

As we have seen in this paper, the wavenumber selection by sources
entails various scenarios for instabilities and chaotic dynamics in
the single-mode patches that are separated by sources and sinks. In
the experiments, there are regimes in parameter space where the
dynamics is reminiscent of what one expects when the mode selected by
the sources becomes convectively or absolutely unstable. Whether the
data are consistent with this scenario has remained unexplored,
however.

We finally note that it has recently become apparent that traveling
waves in convection cells with a free surface which are heated from
the side \cite{garci,nathes,natha}, are intimately related to those
occurring in the heated wire experiments \cite{daviaud}. Sources and
sinks have also been observed in such experiments, but a systematic
study of some of the issues we raise does not appear to have been
undertaken yet. Clearly, both the heated wire experiments and this
system appear to be very suitable setups to study the dynamics of
sources and sinks; in addition, both do show rich dynamical behavior.

\subsubsection{Binary mixtures}\label{sss_binmix}

One of the best studied experimental traveling wave systems is binary
fluid convection \cite{kolss,kolss2,kolss3,binliq1,binliq2}. Since the
bifurcation in this case has been shown to be weakly subcritical
\cite{bensimon}, the description of the behavior
in this system is strictly speaking beyond the scope of the coupled
CGL equations we 
consider. A brief discussion is nevertheless warranted, not only
because some of the behavior of sources and sinks is quite generic, in
that it does not strongly depend on the underlying bifurcation
structure (e.g., sources still form a discrete set according to the
counting arguments), but also because the additional complications of
the binary mixture convection experiments are an interesting subject
for future study.

Kaplan and Steinberg have shown that the transition from localized
traveling wave patterns (pulses) to extended traveling waves is
essentially governed by the convective instability of the edges of the
pulses \cite{kapstein1}\footnote{This is similar to the behavior of
  sources near $\varepsilon^{so}_c$ (section \ref{sss_val}).}.  The
fact that the relevant front velocity is given by linear marginal
stability arguments, suggests that the subcritical character of the
bifurcation is not very strong here.  On the other hand, the
nonadiabatic effects, such as locking, observed in \cite{kapstein2},
point in the other direction, namely that the subcritical nature of
the transition is rather strong. Hence, the importance of the
subcritical effects in these experiments can not be trivially
established.


Kolodner \cite{kolss2} has observed a wide variety of source/sink
behavior. In some cases, there appears to be a stable source/sink pair
where the sink is clearly wider than the source. This of course
contradicts what we typically find (except close to the relaxational
limit --- see section \ref{ss_s0}). This may have to do with the
subcritical nature of the bifurcation, but one should also keep in
mind that in other cases there is evidence that such behavior could
still be a transient, because there are still phase slip events
occurring. E.g., Fig. 5 of \cite{kolss2} shows a notable example of a
case in which the sink is initially wider than the source, but in
which a process clearly involving the fast scales narrows it down, so
that in the end it smaller than the source.

Another interesting state that is encountered in the experiments are
drifting source/sink patterns (see, e.g., Fig.\ 7 of \cite{kolss2}).
The sources here move slowly but with a constant velocity, and are
non-symmetric in that the wavenumbers on either side are different.
However, there is again a one-to-one correspondence between the drift
velocity and the difference in wavenumbers. In \cite{kolss2}, this is
referred to this process as ``Doppler shifting'', to indicate that in
the frame co-moving with it, the drifting source sends out waves with
the same frequency to the left and the right. This is completely
equivalent to the ``phase matching rule'' of \cite{alvarez}. When such
a moving source is present, the sinks are also found to obey the phase
matching rule and so they move with exactly the same drift velocity as
the sources. Clearly, it is still the source that selects the wave
number and hence plays the active role here --- as usual, the sink
motion is essentially determined by the properties of the waves that
come in. A priori, one could imagine that the sources and sinks in the
binary fluid experiments are more prone towards obeying the phase
matching rule due to the subcritical nature of the bifurcations
to traveling waves, but one can find various examples in the
experiments where they do not obey this rule. Obviously, this question
deserves further study.

The fact that Kolodner \cite{kolss2} observes in his Fig.\ 7 a
steadily moving source is not necesarily in contradiction with our
counting arguments, as these do allow for the existence of a discrete
set of $v \neq 0$ sources. In practice, however, for a proper analysis
of such source solutions in the binary fluid experiments it is
probably necessary to include the coupling to the slow concentration
field, as in the work of Riecke and coworkers on traveling pulse
solutions \cite{coupling1,riecke2,riecke3}.

Although several of the experiments of Kolodner have been done at very
small values of $\varepsilon$, there is no visible evidence of the
divergence of the width of any of the sources and sinks. Presumably,
this is due to the subcritical nature of the bifurcation --- in
section \ref{ss_num} we already argued that in this case the width of
neither the sources nor the sinks need to diverge as $\varepsilon
\rightarrow 0$.

In passing, we note that, quite impressively, Kolodner has also been
able to extract the spatial amplitude profiles of his sources and sinks
(Figs. 8, 18 and 21 of \cite{kolss2}). These agree remarkably well
with the profiles 
we obtained numerically using the shooting method described earlier.
Even the characteristic overshoots of the amplitudes near the edges of
sinks are clearly observable in all cases.

In conclusion, although a detailed comparison between the sources and
sinks in binary fluid experiments and those analyzed theoretically
here, is not justified, many qualitative features (multiplicity,
wavenumber selection, etc.) are quite similar. We expect that the
$\varepsilon $ dependence of the width of these structures is very
different in the two cases, due to the subcritical nature of the
bifurcation in binary mixtures and due to the coupling to the slow
concentration field. The latter effect probably also plays an
important role in the drift of the sources.

\subsection{Open problems}

In spite of the fact that we have been able to map out many of the
various possible static and dynamical properties of sources and sinks,
there remains a large number of theoretical issues and open problems
which need to be studied in further detail. This section briefly lists
the ones we consider most important.
  
$\bullet$ {\em Phase matching.} The absence of the coupling of the phases across
a moving sink appears to be one of the main short comings of the
coupled CGL equations.
  
  For the single mode CGL equation, the velocity of sinks
  is determined in terms of the two wavenumbers $q_{N_1}$ and
  $q_{N_2}$ of the incoming modes, without solving for the structure
  of the sinks: $v=(c_1+c_3)(q_{N_1}+q_{N_2})$ \cite{physd}.  This
  follows 
  directly from the requirement that in the frame moving with the
  sink, the frequencies to the left and the right of the sink should be
  equal. Phase slips occur when these frequencies are unequal,
  and in that case the sink is not a ``coherent structure'' (i.e., it has a
  time-dependent spatial profile).
  
  For the sinks in the coupled CGL equations
  (\ref{coupcgl1},\ref{coupcgl2}) that we have studied here, the
  velocity of a moving sink can not be simply given in terms of the
  wavenumbers of the incoming waves --- the velocity is determined
  implicitly by the solution of the ODE's (\ref{al}-\ref{zr}). The
  frequencies to the left and to the right of sinks correspond to two
  different modes, and the coupling between these modes depends only
  on their amplitudes, not on their phase.  Moreover, the phase
  matching as observed empirically in the experiments \cite{alvarez}
  clearly involves the fast scale that has been eliminated to obtain
  the amplitude equations; therefore, such rule can never be implemented
  in the standard coupled CGL Eqs.  (\ref{coupcgl1},\ref{coupcgl2})
  \cite{alvarez}.
  
  The phase matching as observed in the experiments is clearly a
  non-adiabatic effect as it involves both the fast and the slow
  scales. Can this non-adiabatic effect be studied perturbatively, as
  in \cite{nonadi}?  As pointed out to us by Newell, the experimental
  phase matching appears to be the analogue in space-time of what
  happens at grain boundaries in the phase equations in the nonlinear
  regime \cite{newell2}. Does this analogy open up a route towards
  analyzing this effect?
  
$\bullet$ {\em Multiplicities.} In our counting analysis, we have focussed on
  the regime where $|v|\!<\!s_0$, and in particular on the case
  $v\!=\!0$. From the results detailed in the appendix, it follows
  that the flow structure near the fixed points changes when
  $|v|\!>\!s_0$; this implies that the counting arguments allow for
  rapidly moving source and sinks solutions with different
  multiplicities. We do not know whether such solutions actually
  exist. We have not studied this possibility (nor the one associated
  with changes of the fixed point structure related to the critical
  velocity $v_{cN}$) in detail, as we have neither found such coherent
  structure solutions of the ODE's, nor observed any of them in
  numerical simulations of the coupled CGL equations.
  
$\bullet$ {\em Coherent structures.}  When $g_2$ is large enough,
single 
  amplitude coherent structures such as sources, sinks and homoclinic
  holes are often exact solutions of the coupled CGL equations. One of
  the modes corresponds then to the coherent structure, the other mode
  is zero.  To see this, note that solutions of the single CGL
  equation have often a minimum amplitude $a_m$ which is nonzero. As
  long as $\varepsilon_{\it eff} = \varepsilon- g_2 a_m^2$ remains
  negative for the zero mode, this mode is suppressed.  A detailed
  analysis of the behavior of such coherent structures as $g_2$ is
  reduced and the other mode becomes active, remains to be done.
  
  The closely bound source/sink pairs, as shown in Fig. \ref{f7}a can
  be seen as a ``new'' coherent structure of the coupled CGL
  equations.  We 
  note that from a counting point of view, such source-sink pairs
  typically correspond to homoclinic orbits, since they often connect
  the same plane wave state to the left and the right. Irrespective of
  the details of the structure of the corresponding fixed point, one
  needs to satisfy in general one condition to obtain such a
  homoclinic orbit (One can see this easily as follows.  Suppose the
  fixed point has a $n$-dimensional outgoing manifold.  This yields
  $n-1$ degrees of freedom and $n$ conditions, so in general one
  parameter needs to be tuned to obtain a homoclinic orbit). Since we
  have three free parameters, this yields a two-parameter family of
  such sink-source pairs
  
  It would be interesting to investigate whether these homoclinic
  structures are connected to the homoclinic holes, analyzed recently
  for the single CGL equation \cite{homoclons}. It is conceivable that
  upon lowering $g_2$, the suppressed mode will mix in below some
  particular value of $g_2$, so that a homoclinic holes can be
  deformed to coupled sink-source pairs.
  
  A related issue is the study of the cross-over from an array of
  sources and sinks to an (almost) periodically modulated amplitude
  pattern of the type seen in Fig. \ref{f11a} and by Sakaguchi
  \cite{saka1}.
  
$\bullet$ {\em Phase-space and dynamical arguments.} In section \ref{ss_num},
  the existence of a special value $\varepsilon_c^{\it so}$ was
  obtained from what was essentially a dynamical argument. At this
  value of $\varepsilon$, the width of stationary sources, as
  determined by the set of ODE's (\ref{al}-\ref{zr}), was found to
  diverge.  What is the precise connection between the phase-space
  structure of the ODE's and the dynamical argument? This question is
  related to that which arises in the study of nonlinear global modes
, and it is quite possible that the analysis of \cite{couairon} can be
extended to sources as well.
  
$\bullet$ {\em Stability.} A full stability analysis of sources and sinks would
  be welcome, as most of our discussion on their stability is based on
  intuitive arguments. Such an analysis might well detect the
  existence of additional instability mechanisms associated with the
  existence of discrete core modes in much the same way as happened
  for pulses \cite{kaup}.
  
$\bullet$ {\em Breathing.} In section \ref{ss_num}, we noted that interactions
  between local structures and fronts often give rise to an
  oscillatory or ``breathing'' type of dynamics \cite{brand,riecke2}.
  The mechanism through which this happens remain largely unexplored,
  however.
  
  Coullet et al. \cite{coullet2} briefly mention that below
  $\varepsilon_c^{\it so}$, sources are very sensitive to noise. We
  found that the average width of the breathing sources depends weakly
  on the strength of the noise, but have not investigated this issue
  in detail. The dependence on the noise should be clarified further.
  
  Finally, after a long transient, the non stationary sources below
  $\varepsilon^{\it so}_{\it c}$ seem to be only very weakly
  time-dependent, and in some sense ''near'' a stationary source
  solution. Can this idea be made more precise?
  
$\bullet$ {\em Pinning and interactions.} Partly to explain the experimental
  observation of Alvarez {\em et al.} \cite{alvarez}, we have
  conjectured that sources can be pinned to slight inhomogeneities,
  and that if they do, the selected wavenumber will vary with the
  local inhomogeneity.  Moreover, stationary sources are then expected
  to exist below $\varepsilon_{\it c}^{\it so}$ of the homogeneous
  system, in much the same way as boundary conditions can give rise to
  the existence of stable stationary sources below $\varepsilon_{\it
    c}^{\it so}$ \cite{rovin}.  Again, a back-up of these conjectures
  is called for.
  
  As some of our simulations indicate (see Fig. \ref{lastfig}), when
  sources and sinks get close to each other, they attract and
  eventually coalesce (or form a pair) in some characteristic fashion.
  Can this attraction be understood perturbatively?
  
$\bullet$ {\em Bimodal chaos.} One of our key observations is that the
  wavenumber selection induced by the sources allows for a bimodal
  instability for $g_2$ just above 1.  For $g_2$ just below 1, similar
  behavior can be found \cite{biriecke}.  The chaotic dynamics in
  these regimes involves the competition between the two modes in an
  essential way, and apart from \cite{amen,biriecke}, a detailed
  analysis of the dynamics here is lacking.
  
$\bullet$ {\em Subcritical bifurcations.} To what extent can our
arguments be 
  extended to the case of a weakly subcritical bifurcation? As we
  discussed in section \ref{sss_binmix}, this issue is of relevance to
  the experiments on binary mixtures.

Finally, we stress that in most cases we have only shown examples of
the possible types of behavior. A more systematic mapping out of the
phase-space of the coupled CGL equations
(\ref{coupcgl1}-\ref{coupcgl2}) may very well lead to additional
surprises.

\ack We would like to thank Guenter Ahlers, Tomas Bohr, Arnaud
Chiffaudel, Pierre Coullet, Francois Daviaud, Lorenz Kramer, Natalie
Mukolobwiez, Alan Newell, Willem van de Water, and Mingming Wu for
interesting and enlightening discussions. In addition, we wish to
thank Roberto Alvarez for a fruitful collaboration of which this work
is an outgrowth.  MvH acknowledges financial support from the
Netherlands Organization for Scientific Research (NWO), and the EU
under contract nr. ERBFMBICT 972554.

\newpage

\begin{appendix}

\section{Coherent structures framework for the single 
  CGL equation}\label{sss_coh}

\subsection{The flow equations} \label{a.1}
In this appendix, we lay the groundwork for our analysis of the
coupled equations by summarizing and simplifying the main
ingredients of the analysis of \cite{physd}  of the single CGL equation
\begin{equation}\label{cgl}
\partial_t A = \varepsilon A + (1+ i c_1) \partial_x^2 A - (1- i
c_3)|A|^2 A ~.
\end{equation}
Note that if a single mode is present,  the coupled equations reduce
to a single CGL written in the frame moving with the linear group
velocity 
of this mode, {\em not} in the stationary frame.
 
As in Eqs. (\ref{ans2.1}), a  coherent structure is
defined as a solution whose time dependence 
amounts, apart from an overall time-dependent phase factor, to a
uniform translation in time with velocity $v$:
\begin{equation}
  A(x,t):= e^{- i \omega t} \hat{A}(x- vt) = e^{- i \omega t}
  \hat{A}(\xi) \label{ans}~.
\end{equation}
Note that if the coherent structure approaches asymptotically a plane
wave state for $\xi \rightarrow \infty$ or for $\xi \rightarrow
-\infty$, the phase velocity of these waves would equal the
propagation velocity of the coherent structures if $\omega$ would be
$0$. When $\omega \neq 0$, 
these two velocities differ.

For solutions of the form (\ref{ans}), $\partial_t\!=\! -i \omega
- v \partial_\xi$, so when we substitute the Ansatz (\ref{ans}) into
the single CGL equation (\ref{cgl}), we obtain the following ODE:
\begin{equation} \label{ode}
  (-i \omega -v \partial_{\xi} )\hat{A} = \varepsilon \hat{A} + (1+ i
  c_1) \partial_{\xi}^2 \hat{A} - (1-ic_3) |\hat{A}|^2 \hat{A}~.
\end{equation}
Solutions of this ODE correspond to coherent structures of the CGL equation
(\ref{cgl}) and vice-versa \cite{physd}.

To analyze the orbits of the ODE (\ref{ode}), it is useful to rewrite
it as a set of coupled first order ODE's. To do so, it is convenient
to write $A$ in terms of its amplitude and phase
\begin{equation}\label{rep}
  \hat{A}(\xi) := a(\xi) e^{i \phi(\xi)}~,
\end{equation}
where $a$ and $\phi$ are real-valued.  Substituting the representation
(\ref{rep}) into the ODE (\ref{ode}) yields, after some trivial
algebra
\begin{equation} 
\partial_{\xi} a =  \kappa a~, ~~~~~~\label{odea}
\partial_{\xi} \kappa =  {\cal K}(a,q,\kappa)~,~~~~~~\label{odek} \\
\partial_{\xi} q = {\cal Q}(a,q,\kappa)~, \label{odeq}
\end{equation}
where $q$ and $\kappa$ are defined as
\begin{equation} 
  q:=\partial_{\xi} \phi, ~~~ \kappa := (1/a) \partial_{\xi} a~.
\end{equation}  
The fact that there is no fourth equation is due to the fact that the
CGL equation is invariant under a uniform change of the phase of $A$,
so that $\phi$ itself does not enter in the equations.  The functions
${\cal K}$ and ${\cal Q}$ are given by \cite{physd}
\begin{eqnarray}
  {\cal K} :=& \frac{1}{1+c_1^2} \left[ c_1(-\omega-v q) - \varepsilon
  - v \kappa +(1-c_1 c_3) a^2 \right] + q^2-\kappa^2~,\\ 
  {\cal Q} :=& \frac{1}{1+c_1^2} \left[ (-\omega- v q) +
  c_1 (v \kappa + \varepsilon) - (c_1+c_3) a^2\right]-2 \kappa q ~.
\end{eqnarray}
At first sight it may appear somewhat puzzling that we write the
equations in a form containing $\kappa\!=\! \partial_\xi \ln a$
instead of simply $\partial_{\xi} a$. One advantage is that it allows
us to distinguish more clearly between various structures whose
amplitudes vanish exponentially as $\xi \to \pm \infty$ --- these are
then still distinguished by different values of $\kappa$.  Secondly,
the choice of $\kappa$ in favor or 
$\partial_\xi a$ allow us to combine $\kappa$ and $q$ as the real and
imaginary part of the logarithmic derivative of $\hat{A}$: we can
rewrite (\ref{odek}) and (\ref{odeq}) more compactly as
\begin{equation}\label{partz}
\partial_{\xi} z 
= -z^2 +\frac{1}{1+i c_1}\left[ - \varepsilon- i \omega + (1-i c_3)a^2
-v z\right] ~.
\end{equation}
where $z \!:=\! \partial_{\xi} \ln(\hat{A}) \!=\! \kappa + i q$.

The fixed points of the ODE's have, according to (\ref{odea}), either
$a\!=\!0$ or $\kappa\!=\!0$.  The values of $q$ and $\kappa$ for the
$a\!=\!0$ fixed points are related through the dispersion relation of
the linearized equation, or, what amounts to the same, by the equation
obtained by setting the right hand side of (\ref{partz}) equal to zero
and taking $a\!=\!0$. Following \cite{physd} we will refer to these
fixed points as {\em linear fixed points}. We will denote them by
$L_{\pm}$, where the index indicates the sign of $\kappa$.
This means that the behavior
near an $L_+$ fixed point corresponds to a situation in which the
amplitude is growing away from zero to the right, while the behavior
near an $L_-$ fixed point describes the situation in which the
amplitude $a$ decays to zero.

Since a fixed point with $a \neq 0, \kappa\!=\!0$ corresponds to
nonlinear traveling waves,  the corresponding fixed points are refered
to
as {\em nonlinear fixed points}  \cite{physd}. We denote these by
$N_{\pm}$,  where
the index now indicates the sign of the {\em nonlinear group velocity} $s$
of the corresponding traveling wave \cite{physd}. Thus,  since the index of $N$ denotes the sign of the
group velocity, the amplitude near an $N_+$ fixed point can either
grow ($\kappa >0$) or decay ($\kappa <0$)  with increasing $\xi$.

The coherent structures correspond to orbits
which go from one of the fixed points to another one or back to the
original one, and the counting analysis amounts to establishing the
dimensions of the in- and outgoing manifolds of these fixed points. In
combination with the number of free parameters (in this case $v$ and
$\omega$), this yields the multiplicity of orbits connecting these
fixed points, and therefore of the multiplicity of the corresponding
coherent structures.
\subsection{Fixed points and linear flow equations in their
  neighborhood} 
Since there are three flow equations (\ref{odea}), there are three
eigenvalues of the linear flow near each fixed point.  When we perform
the counting analysis for these fixed points we will only need the
signs of the real parts of the three eigenvalues, since these
determine whether the flow along the corresponding eigendirection is
inwards ($-$) or outwards ($+$). We will denote the signs by pluses and
minuses, so that $L_-(+,+,-)$ denotes an $L_-$ fixed point with two
eigenvalues which have a positive real part, and one which has a
negative real part.

From  Eqs.\ (\ref{odea}) and (\ref{partz}), we obtain  as
fixed point equations 
\begin{equation}
a \kappa = 0~, \label{fp1}~~~~~~
(1+i c_1) z^2 + v z + \varepsilon + i \omega - (1+i c_3) a^2 = 0~, \label{fp2}
\end{equation}
where $z\!:=\! \kappa+i q$.
 From (\ref{fp1}) we immediately obtain that fixed points either have
$a\!=\!0$ (linear fixed points denoted as $L$) 
or $a\neq0,\kappa\!=\!0$ (nonlinear fixed points denoted as $N$). 
Defining $\tilde{v} \!:=\! v/(1+ c_1^2)$ and $\tilde{a}\!:=\! a/(1+ c_1^2)$,
the derivative of the flow (\ref{odea}) is given by the matrix:
\begin{equation}
DF = \left( \begin{array}{ccc}
\kappa & a & 0\\
2\tilde{a}(1-c_1 c_3)~~ &
-2 \kappa - \tilde{v}~~&
 2 q  - c_1 \tilde{v}~~   
\\
-2 \tilde{a}(c_1+c_3)~~ &
-2 q + c_1 \tilde{v}~~ &
 -2 \kappa -\tilde{v}~~  
\end{array}\right)~.  \label{defDM}
\end{equation}
Solving the fixed point equations (\ref{fp1},\ref{fp2}) and
calculating the eigenvalues of the matrix DF (\ref{defDM}) yields the
dimensions of the incoming and outgoing manifolds of these fixed
points. Note that according to our convention,  a fixed point with a two-dimensional outgoing and
one-dimensional ingoing manifold is denoted as $(+,+,-)$.

We can restrict the calculations to the case of positive $v$, since
the case of negative $v$ can be found by the left-right
symmetry operation: $ \xi \rightarrow -\xi$, $v \rightarrow -v$, $ z
\rightarrow -z $.

\subsection{The linear fixed points}\label{ss_lfp}
For the linear fixed points $a\!=\!0$, and from (\ref{fp2})
we obtain as fixed-point equation:
\begin{equation}
(1+i c_1)z^2 + v z  + \varepsilon + i \omega =0~,
\label{disp}
\end{equation}
which has as solutions
\begin{equation}
z = \frac{-v \pm \sqrt{v^2 - 4 (1+ i c_1) ( \varepsilon + i \omega)}}
{2(1+ i c_1))}~. \label{sol}
\end{equation}
The linear fixed points come as a pair, and
 the left-right symmetry implies that for $v\!=\!0$,
the eigenvalues of these fixed points are opposite. 

At these fixed points, the eigenvalues are given by
\begin{equation}\label{sq}
\kappa\mbox{  or  }-\tilde{v}-2 \kappa  
\pm i (c_1\tilde{v} -  2 q)~.
\end{equation}
To establish the signs of the real parts of the eigenvalues, we
need to determine the signs of $\kappa$ and $-\tilde{v}-2 \kappa$.

Let us first establish the signs of $\kappa$;
this is  important in establishing whether the evanescent
wave decays to the left ($L_+$) or to the right ($L_-$).
For $v\!=\!0$, the equation (\ref{disp}) is purely quadratic,
and so its solutions come in pairs $\pm (\kappa+i q)$.
By expanding the square-root
(\ref{sq}) for large $v$ one obtains that in this case
$\kappa\!=\!-v$ or $\kappa \!=\! -\varepsilon/v$; for large $v$,
 both $\kappa$'s are negative.
Solving equation (\ref{disp}) we find that $\kappa$ changes sign when
\begin{equation}
q = \pm \sqrt{\varepsilon}~,~~~~~
v = \frac{c_1 \varepsilon - \omega}{\sqrt{\varepsilon}}~.
\end{equation}
For $\varepsilon <0$, these equations have no solutions, and
in that case there always is a $L_-$ and a $L_+$ fixed point. 
For $\varepsilon >0$ and $v< (c_1 \varepsilon -\omega)/\sqrt{\varepsilon}$ 
there also is a $L_-$ and a $L_+$ fixed point; for large $v$, there 
are two $L_-$ fixed points.

To determine the sign of $-\tilde{v}-2 \kappa$
note that from the solution (\ref{sol}), we obtain that
$\kappa \!=\! -\tilde{v}/2 \pm \mbox{Re}({\sqrt{\dots}}/\dots)$.
After some trivial rearranging this yields that
$-\tilde{v}-2 \kappa$ has opposite sign for the pair of $L$ fixed points;
when one of them has two $+$'s, the other has two $-$'s.

In the case that we have a $L_+$ and a $L_-$ fixed point the counting
is as follows.
For the $L_+$ fixed point,  $-\tilde{v}-2 \kappa $ is 
negative since both $v$ and $\kappa$ are positive,
and the eigenvalue structure is then $(+,-,-)$. 
The  $L_-$ fixed point then has one negative eigenvalue $\kappa$,
and two positive eigenvalues coming from the $-\tilde{v}-2 \kappa $.
For large $v$, both $\kappa's$ are negative, and we obtain 
a $L_- (+,+,-)$ and a $L_- ( + ,-,-)$ fixed point. 

In summary, then, the counting for the linear fixed points is as follows:

\begin{eqnarray}
\begin{array}{lll}

&\varepsilon <0~~
&\left.\begin{array}{lll}
\hspace{.1cm}
\mbox{ all }v:~~~ \hspace{.73cm} & L_- (+,+,-) & L_+(+,-,-)~,
 \end{array} \right. \\

&\varepsilon >0~~
&\left\{  \begin{array}{lll}
 v < -v_{cL}:~~~ & L_+ (+,-,-) & L_+ (+,+,+)~,\\
|v|<v_{cL}:~~~ & L_- (+,+,-) & L_+(+,-,-)~,\\
v> v_{cL}:~~~ & L_- (+,+,-) & L_- (-,-,-)~,
         \end{array} \right. \label{lconditions} 
\end{array}
\end{eqnarray} where
$ v_{cL} = |c_1 \varepsilon - \omega| /\sqrt{\varepsilon}$.

\subsection{The nonlinear fixed points}\label{ss_nlfp}

The analysis of the nonlinear fixed points goes along the same lines.
Since the nonlinear fixed point has $\kappa\!=\!0$, $z\!=\!i q$, 
the fixed point equations become:
\begin{equation}
a^2 =  \varepsilon - q^2~,~~~~~~
q^2(c_1 + c_3) - v q - \omega - c_3 \varepsilon = 0~.  \label{a10b}
\end{equation}
which yields
\begin{equation}\label{sq2}
  q=\frac{v\pm\sqrt{v^2 + 4(\omega+ c_3 \varepsilon)(c_1 +
      c_3)}}{2(c_1+c_3)}~.
\label{qnl}
\end{equation}
So the nonlinear fixed points come as a pair.

To obtain the eigenvalues, 
we substitute $\kappa\!=\!0$ in the  (\ref{defDM})
and obtain as a secular equation:
\begin{eqnarray}
(1+ c_1^2) \lambda^3 + 2v \lambda^2 +
&\left[ 2 a^2 (c_1c_3-1) + 
4 q^2(1+c_1^2)-4c_1qv+v^2\right] \lambda + \nonumber\\
&\left[ 4 a^2(c_1 + c_3)q - 2 a^2 v\right] =0~.
\end{eqnarray}
We only need to know the number of solution of
the secular equation that have positive real part, and instead
of solving the equation explicitly,
we can proceed as follows.

For a we cubic equation of the form
\begin{equation}
p_3 \lambda^3 + p_2 \lambda^2 + p_1 \lambda^1 + p_0~, \label{poly}
\end{equation}
where $p_3>0$,  we may read off the signs of the real parts of the
solution to this equation from the following table \cite{physd}:
\begin{eqnarray}\label{psigns}
p_0&>& 0 \left[\begin{array}{llll}
p_2 >  0 , &p_1 p_2 > p_0 p_3: \hspace*{0.5cm} & (-,-,-) \hspace{.1cm} 
&(\mbox{case }i)~,\\
          \mbox{else: } & & (+,+,-)& (\mbox{case }ii)~, \end{array} \right. 
\nonumber \\
p_0&<&0 \left[\begin{array}{llll}
p_2 < 0 , &p_1 p_2 < p_0 p_3:  & (+,+,+)  \hspace{.1cm} &(
\mbox{case }iii)~,\\
          \mbox{else:}  &  & (+,-,-) & (\mbox{case }iv)~.\end{array} \right. 
\end{eqnarray}

According to these rules, there are three combinations of the coefficients 
that we need to now the sign of, being
\begin{eqnarray}
p_0 &=& 4 a^2 q (c_1+c_3) - 2 a^2 v~,\\
p_2&=& 2v~,\\
p_1p_2-p_0p_3 &=& -(1+ c_1^2)\left[4 a^2(c_1+c_3)q - 2 a^2 v\right]
\nonumber
\\ 
&+& 2 v \left[2a^2 \right. (c_1 c_3-1) + 4 q^2 ( 1+ c_1^2) - 4 c_1 q v + v^2
\left . \right] ~.
\end{eqnarray}
As before, we will take $v>0$, which makes $p_2>0$.
 
The sign of $p_0$ is equal to the sign of $2 q ( c_1 + c_3) -v$, 
which according to Eq. (\ref{sq2}) is either $\pm \sqrt {\dots}$.
The group velocity $\partial_q  \omega$ of the
the plane waves corresponding to the $N$ fixed points
 is found from (\ref{a10b}) to be
$2q(c_1+c_3) -v$, which can be rewritten as $p_0/(2a^2)$. 
So, we always have one
$N_-$ fixed point with $p_0<0$ and one $N_+$ fixed point
with $p_0>0$.

When $p_0<0$, since $p_2$ is positive, the fixed point is
$N_- (+,-,-)$ (case $(iv)$). When $p_0>0$, the eigenvalues
depend on the sign of $p_1p_2-p_0p_3$; when it is positive the eigenvalues
are $(-,-,-)$, when it is negative, the eigenvalues are $(+,+,-)$.
Defining $v_{cN}$ as the value of $|v|$ where $p_1p_2-p_0p_3$
changes sign, we obtain for the nonlinear fixed points:

\begin{eqnarray}
\begin{array}{lll}
 v < -v_{cN}: \hspace*{0.4cm}  & N_- (+,+,+) ~~ \mbox{and} & N_+ (+,+,-)~,\\
|v|<v_{cN}: & N_- (+,-,-) ~~ \mbox{and} & N_+(+,+,-)~,\\
v> v_{cN}: & N_- (+,-,-) ~~ \mbox{and} & N_+ (-,-,-)~.
         \end{array}  \label{nconditions}
\end{eqnarray}

Eqs. (\ref{lconditions}) and (\ref{nconditions}) express the
dimensions of the stable and unstable manifolds of the fixed points of
the single CGL equation, and these are the basis for the counting
arguments for coherent structures in this equation \cite{physd}. We
now turn to the extension of these results to the coupled CGL
equations.

\section{Detailed counting for the coupled CGL
  equations}\label{app_coco} 
\subsection{General considerations}
While the counting for the coupled CGL
equations follows unambiguously from that for the single CGL, there
are various nontrivial subtleties in
the extension of those results to the coupled CGL equations that
require careful discussion.

Suppose we want to perform the counting for the
$a_L\!=\!0$, $\kappa_R \!=\! 0$ fixed point, which corresponds to the case in
which only a right-traveling wave is present.  The fixed point
equations that follow from (\ref{zr}) are, up to a change of $v
\rightarrow v-s_0$, equal to the fixed point equation for the
nonlinear fixed points of the single CGL equation, and can be solved
accordingly.  To solve the fixed point equations that follow from
(\ref{zl}), note that $a_R$ is a constant at the fixed point and so
the term $-g_2(1-ic_2)a_R^2$ can be absorbed in the $ -\varepsilon - i
\omega_L$ term.  Since we may choose $\omega_L$ freely, for the
counting analysis we can forget about the $i g_2 c_2 a_R^2$ as we may
think of it as having been absorbed into the frequency.  The sign of
$\varepsilon^L_{\it\!eff\!}$, defined in (\ref{epseffdef}) to be
$\varepsilon^L_{\it\!eff\!}  \! =\! \varepsilon - g_2 a_R^2$ will,
however, be 
important. Likewise, at the other fixed point where $a_R\!=\!\kappa_L\!=\!0 $ the
effective $\varepsilon$ is $\varepsilon^R_{\it\!eff\!}\!=\! \varepsilon - g_2
a_L^2$.

Since the fixed points we are interested in for sources and sinks
always have either $a_L\!=\!0$ or $a_R\!=\!0$, the linearization around them
largely parallels the analysis of the single CGL equation. For, when
we linearize about the $a_L\!=\!0$ fixed point, we do not have to take
into account the variation of $a_R$ in the coupling term and this
allows us, for the counting argument, to absorb these terms into an
effective $\varepsilon$ and redefined $\omega$ as discussed
above. Once this is done, the linear equations for the mode whose
amplitude $a$ vanishes at the fixed point {\em do  not involve the
other mode variables at all}. As a result, the matrix of coefficients
of the linearized equations has a block structure, and most of the
results follow directly from those of the single CGL equation. We will
below demonstrate this explicitly, using a symbolic notation for
various 
terms whose precise expression we do not need explicitly.

If we consider the $6$ variables $a_L$, $\kappa_L$, $q_L$, $a_R$,
$\kappa_R$ and $q_R$ as the elements of a vector $w$, and linearize
the flow equations (\ref{odea}) about a fixed point where one of the
modes is nonzero, we can write the linearized equations in the form
$\dot{w}_i \!=\! \sum_j M_{ij}w_j$, where the $6$$\times$$6$ matrix $M$
has the structure
\begin{equation}
  M = \left( \begin{array}{cccccc} \kappa_L & a_L & 0& 0&0&0\\ "a_L"&
  X&X&"a_R"&0&0\\ "a_L"&X&X&"a_R"&0&0\\ 0&0&0&\kappa_R&a_R&0\\ 
  "a_L"&0&0&"a_R"&X&X\\ "a_L"&0&0&"a_R"&X&X
\end{array}\right)~.  \label{defDF}
\end{equation}
In this expression, all quantities  assume their fixed
point values. Furthermore, $"a_R"$ and $"a_L"$ represent terms that
are linear in $a_R$ or $a_L$, and the $X$ stand for longer expressions
that we do not need at the moment.  At the fixed points, either $a_R$
or $a_L$ is zero, so either the upper-right block is identical to
zero, or the lower-left block is zero. {\em In either case, the
  eigenvalues are simply given by the eigenvalues of the upper-left
  and lower-right block-matrices}.  This implies that for each of the
$3$$\times$$3$ blocks, we can use the results of the counting for a
single CGL equation, provided we take into account that $v$ and
$\varepsilon$ should be replaced by $v\pm s_0$ and
$\varepsilon^L_{\it\!eff\!}$ or $\varepsilon^R_{\it\!eff\!}$ at the appropriate
places!

As discussed in appendix \ref{sss_coh}, the fixed point structure of
the single CGL depends on two 
``critical''  velocities, $v_{cL}$ and $v_{cN}$,
In general, these are different for the two fixed points which the
orbit connects, so there is in principle a large number of possible
regimes, each with their own combination of eigenvalue structures at
the fixed points. An exhaustive list of all possibilities can
be given, but it does not appear to be worthwhile to do so here.  For,
many of the exceptional cases occur for large values of the
propagation velocity $v$ and the relevance of the results for these
solutions of the coupled CGL equations is questionable --- after all,
as we explained before, the counting can at most only demonstrate that
certain solutions might be possible in some of these presumably
somewhat extreme ranges of parameter values, but they by no means
prove the existence of such solutions or their stability or dynamical
relevance. Indeed, as discussed in section \ref{ss_num},  for small
$\varepsilon$ the 
sources are intrinsically dynamical and are not given by the {\em
  coherent} sources as obtained from the ODE's (\ref{al}-\ref{zr}).

For these reasons, our discussion will be guided by the following
observations.  The sinks and sources observed in the heated wire
experiments have velocities that are smaller that the group velocity
\cite{alvarez}\footnote{In the experiments of \cite{alvarez}, it was
  estimated from the data that $s_0 \approx v_{ph}/3$, where $v_{ph}$
  is the phase velocity, while typical sinks had a velocity $v$ which
  could be as small as $v_{ph}/50$.}; this also seems to hold for
other typical experiments with finite linear group velocity $s_0$.
This motivates us to start the discussion by investigating the regime
that the velocity $v$ is smaller than the linear group velocity,
$|v|<s_0$.  The sources are now as sketched in Fig. \ref{f1}a and the
sinks are as in Fig. \ref{f1}c; this restriction already leads to a
tremendous simplification.  Furthermore, we are especially interested
in the case that the two modes suppress each other sufficiently that
the effective $\varepsilon $ of the mode which is suppressed is
negative, i.e., $\varepsilon^{L/R}_{\it\!eff\!} <0$. This requirement is
certainly fulfilled for sufficiently strong cross-coupling.  The
technical simplification of taking $\varepsilon^{L/R}_{\it\!eff\!} <0$
is that in this case the structure of the linear fixed points is
completely independent of the parameters $v$ and $\omega$ --- see Eq.
(\ref{lconditions}). It should be noted, however, that in section
\ref{ss_bim} we will encounter source/sink patterns where
$\varepsilon_{\it eff}$ is positive; these patterns are chaotic.
Also, the {\em anomalous} sources and sinks, mentioned at the end of
section \ref{s_cou}, can in some parameter ranges defy the general
rules 
obtained here (see section \ref{esoteric} of this appendix).
Furthermore, in section \ref{countanom} we will discuss the cases $s_0
<  2 q (c_1 + c_3)$ (i.e., sources and sinks corresponding to those of
Fig. \ref{f1}b and d), and the $s_0 \!=\!0$ limit.

\subsection{Multiplicities of sources and sinks}\label{ss_mul}
We will first perform the analysis starting with the
 restrictions given above.
From Fig. \ref{f1} we can read
off  the building blocks of sources and sinks.
 are.  We refer to the fixed point corresponding to $x\rightarrow
 -\infty (\infty)$ as fixed point 1 (2).  In the coupled CGL equation
 case, we refer to the total group velocity of the nonlinear waves,
 which is given by $ 2 q(c_1 + c_3) +v \pm s_0$ [see Eqs.
 (\ref{sdef1}), (\ref{sdef2})]; since by the substitution $v \rightarrow v
 \pm s_0$ we absorb the $s_0$ in the $v$, the indexes of the $N_-$ and
 $N_+$ fixed points correspond to the nonlinear group velocities in
 the co-moving frame of the coherent structures.  For sinks of the
 type sketched in Fig.  \ref{f1}c, $A_L \!=\!0$ for large negative $x$
 and $A_R \!=\!0$ for large positive $x$.  Consequently, the flow is
 \begin{equation}\label{sinkdef}
 \mbox{from }
 \left\{\begin{array}{ll}
 N_+ &(v-s_0)\\ L_+&(v+s_0)
 \end{array}\right.  \mbox{ to }
 \left\{\begin{array}{ll}
 L_- &(v-s_0)\\ N_-&(v+s_0)
 \end{array}\right. ~.
 \end{equation}

 For sources of the type sketched in Fig. \ref{f1}a, $A_R\!=\!0$ for
 large negative $x$ and $A_L \!=\!0$ for large positive $x$.
 Consequently, the flow is
 \begin{equation}\label{sourcedef}
 \mbox{from }
 \left\{\begin{array}{ll}
 N_- &(v+s_0)\\ L_+&(v-s_0)
 \end{array}\right.  \mbox{ to }
 \left\{\begin{array}{ll}
 L_- &(v+s_0)\\ N_+&(v-s_0)
 \end{array}\right. ~.
 \end{equation}
 As in appendix \ref{sss_coh}, we will denote the real parts of the three eigenvalues of
 the fixed points by a string of plus or minus signs; e.g. $(+,-,-)$.
 
 For $\varepsilon_{\it\!eff}\!<\!0$ and arbitrary velocities, we
 obtain for the $L$ fixed points (see Eqs. (\ref{lconditions})):
 \begin{equation}\label{l1}
 L_- (+,+,-)~,~~~~~ L_+ (+,-,-)~.
 \end{equation}
 For now we assume that $|v|<s_0$, $v-s_0 <0$ and $v+s_0>0$. This
 yields, according to (\ref{nconditions}) for the $N$ fixed points:
 \begin{equation}\label{l2}
 N_- (+,-,-)~,~~~~~~ N_+ (+,+,-)~.
 \end{equation}

 For sources we find that the combined $(N_-,L_+)$ fixed point 1 has a
 two-dimensional outgoing 
 manifold, which yields one free parameter. We can think of this
 parameter as a coordinate parameterizing the ``directions'' on the
 unstable manifold\footnote{Note that a one-dimensional manifold
   yields no free parameters other than the one associated with the trivial translation symmetry of the solution, and, in general, a $p$-dimensional
   outgoing manifold yields $p-1$ nontrivial free parameters}.  Now, the only
 other freedom we have for the trajectories out of fixed point 1 is
 associated with the freedom to view $v$, $\omega_L$ and $\omega_R$ as
 parameters in the flow equations that we can freely vary.  This
 yields a total of four free parameters.  Fixed point 2 (a $(N_+,L_-)$
 combination) has, according
 to Eqs.  (\ref{sourcedef}-\ref{l2}), a four-dimensional outgoing
 manifold.  An orbit starting from fixed point 1 has to be
 ``perpendicular'' to this manifold in order to flow to fixed point 2;
 this yields four conditions.  Assuming that these conditions can be
 obeyed for some values of the free parameters, it is clear that as
 long as there are no accidental degeneracies, we expect that there is
 at most only a discrete set of solutions possible --- in other words,
 solutions will be found for sets of isolated values of the angle,
 $v$, $\omega_L$ and $\omega_R$. One refers to this as a discrete set
 of sources.

 When we fix $v\!=\!0$, there is the following symmetry that we have
 to take into account: $\xi\rightarrow -\xi, z_L\leftrightarrow -z_R,
 a_L \leftrightarrow a_R$. Furthermore, this left-right symmetry
 yields that we should take $\omega_L=\omega_R$, so, in comparison to
 the general case, we have two free parameters less.  When the
 outgoing manifold of fixed point 1 intersects the hyper-plane
 $z_L\!=\!-z_R, a_L\!=\!a_R$, this yields, by symmetry, a heteroclinic
 orbit to fixed point 2.  Therefore we only need to intersect the
 hyper-plane to obtain a heteroclinic orbit, which yields two
 conditions (instead of four in the general case). For the sources we
 have now two conditions and two free parameters; and this yields a discrete set of $v\!=\!0$ sources.
 In other words, within the discrete set of sources we generically
 expect there to be a $v\!=\!0$ source solution.
 
 For a sink we obtain, combining (\ref{sinkdef}, \ref{l1}) and
 (\ref{l2}), that fixed point 1 (a $(N_+,L_+)$ combination) has a
 three-dimensional outgoing 
 manifold, which yields two free parameters, while fixed point 2 (a
 $(N_-,L_-)$ combination) has a
 three-dimensional outgoing manifold, which yields three conditions to
 be satisfied. Together with the three free parameters $v, \omega_L$
 and $\omega_R$, this yields a two-parameter family of sinks.

\subsection{The role of $\varepsilon$}\label{app_eps}

When discussing the counting for the single CGL equation, the value of
$\varepsilon$ is uniquely determined. In the coupled equations however, 
one needs to work with the {\em effective} value of $\varepsilon$ when
studying the linear fixed points, since the growth of the
linear modes are determined by renormalized values of $\varepsilon$
 which are given by
$ \varepsilon_{\it eff,L}  =  \varepsilon-g_2 a_R^2$, 
$\varepsilon_{\it eff,R}  =  \varepsilon-g_2 a_L^2$
for the left- and right-traveling modes respectively [see Eq. (\ref{epseffdef})]. While the
inclusion of the sign structure of the linear fixed points for
positive values of $\varepsilon$ may have seemed somewhat superfluous
for the {\em single} CGL equation,  in the case of the
coupled equations this is relevant. In the analysis in sections
\ref{roleofv}--\ref{countanom} we assume that both effective values of $\varepsilon$
are negative. Some comments on the counting for positive values of
$\varepsilon_{\it eff}$ are given in section \ref{esoteric}.

\subsection{The role of the coherent structure velocity $v$}\label{roleofv}

In the counting for the single CGL equation, we were able to remove the 
group velocity term $\sim s_0$ by means of a Galilean transformation to
the comoving frame. In the coupled equations this is not possible, however, and
we need to incorporate the $s_0$-terms when studying the fixed point structure.

In particular, when translating the result for the single CGL into coupled
CGL variables, we need to make the following replacements where $v$ is
concerned 

\begin{eqnarray}
\mbox{For the $a_R$ mode} & : & v \rightarrow v-s_0 \equiv v_R~,\\
\mbox{For the $a_L$ mode} & : & v \rightarrow v+s_0 \equiv v_L~.
\end{eqnarray}

Just like the possible occurrence of positive values of $\varepsilon$
could possibly affect the linear fixed points, this may well affect
the nonlinear fixed points. In the single CGL equation we were allowed
to take $v \ge 0$, but we can no longer do this in the coupled case.
Let us focus on the case $v=0$, i.e, consider stationary coherent
structures. Since $s_0$ is by definition positive, the $a_L$ mode has
$v_L=s_0 >0$, while the $a_R$ has $v_R=-s_0<0$. The statement that we
can alway take $v>0$ therefore no longer holds here, and we need to
exercise caution when evaluating the nonlinear fixed points as well.
In particular, {\em moving sources} ($v>0$) with $|v_R| >v_{cN}$ or
$v_L> v_{cN}$ can have different multiplicities than the stationary
one with $v=0$.

In the formulas for the counting, one should keep in mind that the
linear group velocities have opposite signs for the left- and right
moving modes: this is also apparent from Eqs.
(\ref{sdef1},\ref{sdef2}), where we defined
$s_{0,R}\!=\!s_0\!=\!-s_{0,L}$, so that we may write the nonlinear
group velocities as
\begin{equation}\label{gvels}
s_R = s_{0,R} + 2 q_R(c_1+c_3)~,~~
s_L = s_{0,L} + 2 q_L(c_1+c_3)~.
\end{equation}

\subsection{Normal sources always come in discrete sets}

In this section, we show that it is not possible for normal stationary sources,
i.e., sources whose $s$ and $s_0$ have the same sign, and for whom
$\varepsilon_{\it eff}<0$ for the linear modes, to come in families. The
flow for a normal source is

\begin{equation}
\mbox{from }
\left\{\begin{array}{ll}
A_L : N_- &(v+s_0)\\ A_R : L_+&(v-s_0)
\end{array}\right.  \mbox{ to }
\left\{\begin{array}{ll}
A_L : L_- &(v+s_0)\\ A_R : N_+&(v-s_0)
\end{array}\right. ~.
\end{equation} 
According to the counting, we have for the $N_-(v+s_0)$ fixed point on the left
that (we take $v=0$)
\begin{eqnarray}
p_0 & = & 4 a_L^2 q_L (c_1+c_3)-2 a_L^2 v_L
=  2 a_L^2 [-s_0+2 q_L(c_1+c_3)]~,\nonumber\\
 & = & 2 a_L^2 s_L < 0 ~,
\end{eqnarray}
because for a normal source $s_L$ has the same sign as $s_{0,L}$. Furthermore
we have
\begin{equation}
p_2 = 2v_L = 2s_0 > 0~.
\end{equation}

This implies, according to Eq. (\ref{psigns}), that the sign structure
of the left fixed point is a $(N_-(+,-,-), L_+(+,-,-))$ combination,
independent of 
the selected wavenumber of the nonlinear mode and the sign of the
combination $p_1p_2-p_0p_3$.  The dimension of the outgoing manifold is
therefore always equal to 2, yielding one free parameter.  For the
right fixed point, a completely similar argument yields an
$(N_+(+,+,-), L_-(+,+,-))$ fixed point, again independent of the selected wavenumber or ${\it
  sgn}[p_1p_2-p_0p_3]$. We therefore have to satisfy 4 conditions at this
fixed point.

Combining this with the free parameters we already had and the
additional symmetry at $v=0$ we find that the sources {\em always}
come in discrete sets, independent of the selected wavenumbers and the
parameters.

\subsection{Counting for anomalous $v=0$ sources}\label{countanom}

When the signs of the linear group velocity $s_0$ and the nonlinear group velocity $s$ are opposite, we are dealing with anomalous sources. This section 
investigates the consequences this has for the counting of such sources.

For an anomalous source, cf. Fig.\ref{f1}b, the flow is (again we only
consider $\varepsilon_{\it eff}<0$ for the linear modes)
\begin{equation}
\mbox{from }
\left\{\begin{array}{ll}
A_L : L_+ &(v+s_0)\\ A_R : N_-&(v-s_0)
\end{array}\right.  \mbox{ to }
\left\{\begin{array}{ll}
A_L : N_+ &(v+s_0)\\ A_R : L_-&(v-s_0)
\end{array}\right. ~,
\end{equation} 
which yields for the nonlinear fixed point on the left
\begin{eqnarray}
p_0 & = & 4 a_R^2 q_R(c_1+c_3) - 2a_R^2v_R
 = 2 a_R^2[ s_0 + 2 q_R(c_1+c_3)]\nonumber~, \\
 & = & 2 a_R^2 s_R <0~.
\end{eqnarray}
where ${\rm sgn}[s_R]=-{\rm sgn}[s_{0,R}]$. Furthermore
\begin{equation}
p_2 = 2v_R = -2s_0<0~,
\end{equation}

so that both $p_0$ and $p_2$ are negative, which implies that,
according to Eq.(\ref{psigns}), the sign structure of the $N_-$ fixed
point depends on ${\rm sgn}[p_1p_2-p_0p_3]$. In particular, when
$p_1p_2-p_0p_3$ is negative it is $N_-(+,+,+)$, and if it is positive
it is $N_-(+,-,-)$. If $p_1p_2-p_0p_3<0$, we can perform a similar
calculation for the right fixed point, and we find that the counting
then yields a 2-parameter family of anomalous sinks. If the expression
is positive, however, we find that the anomalous sources also come in
a discrete set.

The sign of this expression depends, for any given set of
coefficients, on the selected wavenumber $q_{\it sel}$ of the
nonlinear mode, and therefore the wavenumber selection mechanism will
determine whether we can actually get to a regime where sources come
as a family.  In
practice, we have not found any examples where this happens. This 
suggests to us that the possible regions of parameters space where
this might happen, are small.

\subsection{Counting for anomalous structures with $\varepsilon_{\it
    eff}>0$ for the suppressed mode }\label{esoteric} \label{eso_eeff}
As mentioned before, another situation that can change the counting is
realized when the suppression of the effective $\varepsilon$ by the
nonlinear mode is not sufficiently large at the linear fixed points,
so that $\varepsilon_{\it eff}>0$.  If we restrict ourselves to the
$v=0$ case, Eq. (\ref{lconditions}) tell us that the counting may
indeed change when in addition $|s_0|>v_{cL}$. This implies that the
multiplicity of sources and sinks changes dramatically under these
circumstances.  An insufficient suppression may happen in particular
when $g_2$ is only slightly bigger than 1, while the selected
wavenumber is large enough to lower the asymptotic value of the
nonlinear amplitude significantly below its maximal value $\sqrt
\varepsilon$. The zero mode then no longer stays suppressed; instead,
it starts to grow, and we then typically get chaotic dynamics, see,
e.g., section \ref{ss_bim}.  For this reason, we confine ourselves to
a few brief observations concerning the $v=0$ case.

 For $v=0$ and $\varepsilon_{\it eff}>0$, we can, according to Eq.
 (\ref{lconditions}), have a $L_-(---)$ fixed point of the $A_L$ mode when
 $s_0\!>\!v_{cL}$. The $A_R$ mode then has a $L_+(+,+,+)$ fixed point.
 Since the index of $L$ denotes the sign of the asymptotic value of
 $\kappa$, with these fixed points we could in principle build a
 2-parameter family of stationary sources, provided $s$ and $s_0$ have
 the same sign in the nonlinear region; otherwise the structures would
 be anomalous sinks.

 Although we have not pursued the possible properties of such sources,
 we expect almost all members of this double family to be unstable.
 The reason for this is that when $\varepsilon_{\it eff}$ is positive,
 the dynamics of the leading edge of the suppressed mode is
 essentially like that of a front propagating into an unstable state.
 As is well known \cite{physd}, in that case there is also a 2-parameter
 family of fronts in the CGL equation, but almost all of them are
 dynamically irrelevant.

\section{Asymptotic behavior of sinks for $\varepsilon \downarrow 0$}\label{app_epszero}

In this appendix, we will discuss the scaling of the width of sinks in the small-$\varepsilon$ limit.

We will assume  that in the domain to the left of the
sink, the $A_R$-mode is suppressed, i.e., $\varepsilon^L_{\it\!eff\!}<0$
(likewise to the right of the sink).
As will be discussed in section \ref{ss_bim} below, we may get
anomalous behavior when $\varepsilon_{\it\!eff\!,L}>0$, which can occur
when $g_2 a_R^2 < \varepsilon$; in that case the $A_L$ mode is (weakly)
unstable and various types of disordered behavior occur.

Assuming $\varepsilon^L_{\it\!eff\!}$ to be negative
to the left of a sink, the
amplitude of the left-traveling mode 
grows exponentially for increasing $\xi$ as $|A_L| (\xi) \sim e^{\kappa_L^+ \xi}$.
The spatial growth rate $\kappa_L$ is given, by definition, by the
value of $\kappa$ at the linear fixed point. According to Eq.
(\ref{sol}), one finds for $z_L \!=\!
\kappa_L + iq_L$:
\begin{equation}
  z_L= \frac{-(v+s_0) \pm \sqrt{(v+s_0)^2-4(1+ic_1)(\varepsilon_{\it
        eff,L} +i\omega)}}{2(1+ic_1)}~,
\end{equation}
where we have used the fact that for the left-traveling mode, $v$ as used in
the appendix is replaced by $v+s_0$, and
$\varepsilon_{\it\!eff\!,L}\!=\!\varepsilon-g_2 a_R^2$. If we expand the
square-root in the small $\varepsilon$ regime, where $\omega$ also tends to
zero, we obtain
\begin{equation}
  z_L \approx \frac{-(v+s_0)}{2(1+ic_1)} \pm \frac{(v+s_0)}{2(1+ic_1)}
  \left[ 1- \frac{2(1+ic_1)(\varepsilon_{\it\!eff\!,L}+i
      \omega)}{(v+s_0)^2} \right]~.
\end{equation}
Since $\varepsilon_{\it\!eff\!,L}$ is negative, and
of order $\varepsilon$,
the root $z_L^+$ with the positive real part is therefore
\begin{equation}
z_L^+ \approx \frac{-\varepsilon_{\it\!eff\!,L}-i\omega}{(v+s_0)}~,
\end{equation}
so that $\kappa_L^+$ scales with $\varepsilon$ as 
\begin{equation}
\kappa_L^+ = {\tt Re}[z_L^+] \sim \varepsilon~.
\end{equation}
In order for the exponent in $|A_L(\xi)| \sim e^{\kappa^+_L \xi}$  to
be of order unity, $\xi 
\sim {\kappa_L^+}^{-1}\sim \varepsilon^{-1}$, which shows that
the width of the
sinks will asymptotically scale as $\varepsilon^{-1}$ for small
$\varepsilon$.

\end{appendix}

\newpage
\begin{figure}\label{figuress}
 \epsfxsize=1.2\hsize
 \mbox{\hspace*{-.1 \hsize}
}
\caption[]
{  Schematic representations of the various coherent structures that we
  will encounter in this paper. The amplitude of the left (right)
  traveling waves is indicated by a thick (thin) curve, while the
  linear group velocity and total group velocity are denoted by $s_0$
  and $s$ respectively, and their direction is indicated by arrows.
  (a) and (b) are, in our definition, both sources, since the
  nonlinear group velocity $s$ points outward; the majority of cases
  that we will encounter will be of type (a). Similarly, (c) and (d)
  both represent sinks.  Finally, one may in principal encounter
  structures that are neither sources nor sinks. We never have
  observed a structure of the form shown in (e) in our simulations,
  but structures like shown in (f) occur quite generally in the
  chaotic regimes. The dotted curve for the $A_R$ mode indicates that
  we can have many different possibilities here, including the case
  were $A_R\!=\!0$; in that case a description in terms of a single
  CGL equation suffices. Note that figure (f) does not exhaust all
  possibilities which are essentially single-mode structures. E.g., in
  our simulations presented in Fig. \ref{fdouble}, we encounter a case
  where in between a source of type (a) and one of type (b) there is a
  single-mode sink, for which $s$ points inwards.}\label{f1}
\end{figure}

\begin{figure}
 \epsfxsize=1.\hsize
 \mbox{\hspace*{-. \hsize}
}
\caption[]
{(a) Space-time plot showing the evolution of the amplitudes $|A_L|$
  and $|A_R|$ in the CGL equations starting from random initial
  conditions. The coefficients were chosen as
  $c_1\!=\!0.6,c_2\!=\!0.0,c_3\!=\!0.4,s_0\!=\!0.4,g_2\!=\!2$ and
  $\varepsilon \!=\!1$. The grey shading is such that patches of $A_R$
  mode are light and the $A_L$ mode are dark.
  (b) Amplitude profiles of the final state of (a), showing a typical
  sink/source pattern.  (c) Comparison between the source obtained
  from direct simulations of the CGL equations as shown in (b)
  (squares) and profiles obtained by shooting in the ODE's
  (\ref{al}-\ref{zr}) (full curves). (d) Similar comparison, now for
  the wavenumber profiles. In (c) and (d), the thick (thin) curves
  correspond to the left (right) traveling mode.}\label{f3}
\end{figure}

\begin{figure}
  \vspace{0mm} \epsfxsize=1.0\hsize \mbox{\hspace*{-.0 \hsize}
}
\caption[]
{(a,b) Space-time plots showing $|A_R|$ (a) and $|A_L|$ (b) in a
  situation in which there are two different sources present.
  Coefficients in this simulation are $c_1\!=\!3.0$, $c_2\!=\!0$,
  $c_3\!=\!0.75$, $g_2\!=\!2.0$, $s_0\!=\!0.2$ and
  $\varepsilon\!=\!1.0$.  Initial conditions were chosen such that a
  well-separated source-source pair emerges, and a short transient has
  been removed.  The source at $x\approx 730$ is anomalous, i.e., its
  linear and nonlinear group velocity $s_0$ and $s$ have opposite
  signs.  Sandwiched between the sources is a single-mode sink,
  traveling in the direction of the anomalous source; this sink is
  visible in (b).  (c) Snapshot of the amplitude profiles of the two
  sources and the single mode sink at the end of the simulation shown
  in (a-b). (d) The wavenumber profiles of the two sources in their
  final state.  Note that when the modulus goes to zero, the
  wavenumber is no longer well-defined; we can only obtain $q$ up to a
  finite distance from the sources. The selected wavenumber emitted by
  the anomalous source is $q_{sel}=0.387$, while the wavenumber
  emitted by the ordinary source is $q_{sel}=0.341$.  The velocity of
  the sink in between agrees with the velocity that follows from a
  phase-matching rule, i.e., the requirement that the phase difference
  across the sink remains constant. In (c) and (d), thick (thin)
  curves correspond to left (right) traveling modes.}
\label{fdouble}
\end{figure}

\begin{figure}
 \epsfxsize=.99\hsize
 \mbox{\hspace*{-.0 \hsize}
}
\caption{(a) Sketch of a wide source, indicating the 
  competition between the linear group velocity $s_0$ and the front
  velocity $v^*$.  (b) Width of coherent sources as obtained by
  shooting, for $c_1\!=\!c_3\!=\!0.5, c_2\!=\!0,g_2\!=\!2$ and
  $s_0\!=\!1$. (c) Example of dynamical source for same values of the
  coefficients and $\varepsilon=0.15$. The order parameter shown here
  is the sum of the amplitudes $|A_L|$ and $|A_R|$, and the total time
  shown here is 1000. (d) Average inverse width of sources for the
  same coefficients as (b) as a function of $\varepsilon$. The thick
  curve corresponds to the coherent sources as shown in (b). For
  $\varepsilon$ close to and below $\varepsilon_c^{\it so}=0.2$, there
  is a crossover to dynamical behavior.  The inset shows the region around
  $\varepsilon=0$, where the average width roughly scales as
  $\varepsilon^{-1}$.}\label{wide}
\end{figure}

\begin{figure}
\vspace{0mm} \epsfxsize=.98\hsize \mbox{\hspace*{.02 \hsize}
}
\caption[]
{The width of stationary sinks obtained from the ODE's
  (\ref{al},\ref{zr}) as a function of $\varepsilon$, for
  $c_1\!=\!0.6$, $c_3\!=\!0.4$, $c_2\!=\!0$, $s_0\!=\!0.4$,
  $g_0\!=\!1$ and $g_2\!=\!2$. (a) Example of the stationary sink
  which has an incoming wavenumber corresponding to the wavenumber
  that is selected by the sources, for $\varepsilon\!=\!0.5$. (b) {\it
    Idem}, now for $\varepsilon\!=\!0.05$.  Notice the differences in
  scale between (a) and (b). These two sinks  are not related by
  simple scale transformations; this illustrates again the absence of
  uniform $\varepsilon$ scaling of the coupled CGL equations.  (c) As
  $\varepsilon$ is decreased, the sink width initially roughly increases as
  $\varepsilon^{-1/2}$. When $\varepsilon$ becomes sufficiently small,
  the group-velocity terms dominate over the diffusive/dispersive
  terms, and the sink-width is seen to obey an asymptotic
  $\varepsilon^{-1}$ scaling (see (d) for a blowup around $\varepsilon
  \!=\!0$. The straight line indicates the analytic result for the
  50\% width as obtained from Eq. (\ref{solution}), i.e.
  $\mbox{width}^{-1}\!=\!5 ~ \varepsilon/(2\ln{3})$.}\label{sinkwidth}
\end{figure}

\begin{figure}
\vspace{0mm} \epsfxsize=1.1\hsize \mbox{\hspace*{-.12 \hsize}
}
\caption[]
{ Frequency $\omega$,
  corresponding selected wavenumber $q_{sel}$ and perturbation
  velocity $v^*_{BF}$ as a function of $c_2$, for
  $\varepsilon\!=\!1,c_1\!=\!c_3\!=\!0.9, s_0\!=\!0.1$ and $g_2\!=\!2$.
  For $c_2 <-0.25$, $v^*_{BF} <0$, and perturbations in the
  right-flank of the source propagate to the left, so that the waves are
  absolutely unstable. }\label{f5}
\end{figure}

\begin{figure}
  \vspace{0mm} \epsfxsize=1.0\hsize \mbox{\hspace*{-.0 \hsize}
}
\caption[]
{Source/sink patterns with absolutely unstable selected wavenumbers
  for the same coefficients as in Fig. \ref{f5} and various values of
  $c_2$. (a) $c_2\!=\!-0.3$, (b) $c_2\!=\!-0.4$, (c) $c_2\!=\!-0.6$,
  (d) $c_2\!=\!-0.8$. For more information see text.}\label{f7}
\end{figure}

\begin{figure}
  \vspace{0mm} \epsfxsize=1.0\hsize \mbox{\hspace*{-.0 \hsize}
}
\caption[]
{Two more examples of nontrivial dynamics in the absolutely unstable
  case.  Both cases: $c_1\!=\!c_3\!=\!0.9, c_2\!=\!-2.6,g_2\!=\!2$,
  and a transient of $10^4$ is deleted. (a-b): $s_0\!=\!0.1$. Here the
  periodic states are quite dominant. It appears that these states
  themselves are prone to drifting and slow dynamics. (b) Snapshots of
  $|A_L|$ (thick curve) and $|A_R|$ (thin curve) in the final state.
  Obviously, the two modes, although disordered, suppress each other
  completely. (c-d) Here we have increased $s_0$ to $0.2$.  The plane
  waves are still absolutely unstable, and the dynamics is disordered,
  but much less than in case (a-b).}\label{f8}
\end{figure}

\begin{figure}
  \vspace{0mm} \epsfxsize=1.0 \hsize \mbox{\hspace*{-.0\hsize}
}
\caption[]
{Two examples of bimodal chaos. (a) and (c) show space time plots, and
  the grey shading is the same as before. Both simulations started
  from random initial conditions, and a transient of $t\!=\!10^4$ has
  been deleted from these pictures. For a detailed description, see
  text.  Note that the final states of runs (a) and (c), depicted in
  (b) and (d), clearly show that the two modes no longer suppress each
  other completely.}
  \label{f9}
\end{figure}

\begin{figure}
  \vspace{0mm} \epsfxsize=1.0\hsize \mbox{\hspace*{-.0 \hsize}
    \vspace{-0mm} 
}\caption[]
{Two examples of the combination of phase slips and a value of $g_2$
  just above 1. The coefficients are
  $c_1\!=\!1,c_3\!=\!1.4,c_2\!=\!1,\varepsilon\!=\!1, s_0\!=\!0.5$.
  Grey shading as before (right (left) traveling waves are light
  (dark)). In (a), $g_2\!=\!1.05$, while in (b) $g_2\!=\!1.2$.}
\label{f10}
\end{figure}

\begin{figure}
  \vspace{0mm} \epsfxsize=1.0\hsize \mbox{\hspace*{-.0 \hsize}
}
\caption[]
{Space-time plots in the coupled-intermittent regime. To be able to
  show both the dynamics in the $A_L$ and $A_R$ mode, the grey shading
  corresponds to $2 |A_R|+|A_L|$. This yields that right traveling
  patches are brighter in shade than left-traveling patches.  (a)
  $c_1\!=\!0.2, c_2\!=\!c_3\!=\!2, \varepsilon\!=\!1,s_0\!=\!0$ and
  $g_2\!=\!1.2$.  (b) Same coefficients as (a), except for
  $g_2\!=\!1.5$.  (c)
  $c_1\!=\!0.6,c_3\!=\!1.4,c_2\!=\!1,\varepsilon\!=\!1,s_0\!=\!0.1$
  and $g_2\!=\!2$.  (d) Same coefficients as (c), except for
  $c_2\!=\!0$.  For as more detailed description see text.}
\label{fsti}
\end{figure}

\begin{figure}
  \vspace{0mm} \epsfxsize=1.0\hsize \mbox{\hspace*{-.0 \hsize}
}
\caption[]
{Four space-time plots, showing the transition from standing waves to
  disordered patterns, for
  $g_2\!=\!1.1,c_1\!=\!0.9,c_3\!=\!2,s_0\!=\!-0.1,\varepsilon\!=\!1$,
  and (a) $c_2\!=\! -0.72$, (b) $c_2\!=\!-0.71$, (c) $c_2\!=\!-0.5$,
  (d) $c_2\!=\!0$. See text.}
\label{f11a}
\end{figure}

\begin{figure}
  \vspace{0mm} \epsfxsize=1.0\hsize \mbox{\hspace*{-.0 \hsize}
}
\caption[]
{Four space-time plots for the same coefficients as 
in Fig. \ref{f11a}, but now for positive values of $c_2$.
(a) $c_2\!=\!0.8$, (b) $c_2\!=\!0.9$, (c) $c_2\!=\!0.95$, (d) $c_2\!=\!1$.
}
\label{f11b}
\end{figure}

\begin{figure}
  \vspace{0mm} \epsfxsize=1.0\hsize \mbox{\hspace*{-.0 \hsize}
}
\caption[]
{(a) Space-time plot of $|A_R|$ illustrating the interaction between
  sources and sinks. The runs started from random initial conditions,
  and the coefficients where chosen as $c_1\!=\!0.6$, $c_3\!=\!0.4$,
  $c_2\!=\!0$, $g_2\!=\!2.0$, $s_0\!=\!0.4$ and at $\varepsilon \!=\!
  0.07$.  Note that $\varepsilon$ is well above the critical value
  $\varepsilon_c^{so}=0.029$, and the sources are stable. Hence, any
  movement of the coherent structures is solely due to their
  interactions. Note that in the final stage of an annihilation event,
  the source moves most, while the sink stays almost put. Note also
  the similarity to Fig. 24 of \cite{kolss2}.  (b) Hidden line plot of
  $|A_L|$ showing the annihilation process in detail.}
\label{lastfig}
\end{figure}

\end{document}